\pdfoutput=1
\documentclass[sigconf]{acmart}

\AtBeginDocument{%
  }

\usepackage{comment}
\usepackage{capt-of}

\usepackage{graphicx}
\graphicspath{{./}}
\usepackage{tikz}
\usetikzlibrary{arrows.meta,calc,positioning,decorations.pathmorphing,
                decorations.markings,backgrounds,fit,shapes.geometric}

\makeatletter
\let\mdPrevRM\rmdefault
\makeatother
\usepackage{XCharter}
\makeatletter

\renewcommand{\rmdefault}{\mdPrevRM}
\makeatother

\definecolor{mdInk}{HTML}{26231E}      
\definecolor{mdSpeech}{HTML}{32517D}   
\definecolor{mdDeixis}{HTML}{C2452D}   
\definecolor{mdGesture}{HTML}{B8862B}  
\definecolor{mdGrey}{HTML}{6E675E}     
\definecolor{mdRule}{HTML}{CFC8BB}     
\definecolor{mdSoft}{HTML}{F1EDE3}     
\definecolor{mdSofter}{HTML}{F8F5EE}   

\newcommand{\mdFa}{\rmfamily\scshape\fontsize{9.4}{10.8}\selectfont}
\newcommand{\mdFb}{\sffamily\fontsize{7.5}{8.8}\selectfont}
\newcommand{\mdFc}{\sffamily\itshape\fontsize{7.0}{8.2}\selectfont}
\newcommand{\mdFm}{\ttfamily\fontsize{7.2}{8.4}\selectfont}

\tikzset{
  md/.style           = {line width=0.75pt, line cap=round, line join=round},
  mdhair/.style       = {line width=0.35pt, draw=mdRule},
  mdarrow/.style      = {line width=1.0pt, draw=mdInk, -{Stealth[length=4.4pt,width=3.8pt]}},
  mdflow/.style       = {line width=0.8pt, draw=mdGrey,
                         -{Stealth[length=4.0pt,width=3.4pt]}},
  mdpanel/.style      = {draw=mdRule, line width=0.4pt, fill=white},
  mdcard/.style       = {draw=mdRule, line width=0.4pt, fill=mdSofter},
  mdchip/.style       = {draw=mdRule, line width=0.4pt, fill=white,
                         inner sep=2.2pt, align=center,
                         font=\mdFb, text=mdInk},
  mdtitle/.style      = {font=\mdFa, text=mdInk, align=left},
  mdlabel/.style      = {font=\mdFb, text=mdInk, align=left},
  mdnote/.style       = {font=\mdFc, text=mdGrey, align=left},
  mdnotec/.style      = {font=\mdFc, text=mdGrey, align=center},
  mdmono/.style       = {font=\mdFm, text=mdInk, align=left},
  mdbrush/.style      = {line width=1.9pt, line cap=round, line join=round,
                         draw=mdGesture},
  mdbrushink/.style   = {line width=1.6pt, line cap=round, line join=round,
                         draw=mdInk},
  mdwaver/.style      = {decorate, decoration={random steps,
                         segment length=3.4mm, amplitude=0.22pt}},
}

\definecolor{nfInk}{HTML}{262A27}
\definecolor{nfBlue}{HTML}{234C63}
\definecolor{nfRed}{HTML}{AA3E2D}
\definecolor{nfAmber}{HTML}{80601F}
\definecolor{nfTeal}{HTML}{426653}
\definecolor{nfGrey}{HTML}{5D6258}
\definecolor{nfLine}{HTML}{B4BCB0}
\definecolor{nfBg}{HTML}{F7F8F6}
\definecolor{nfStone}{HTML}{E5E7E0}
\definecolor{nfPlane}{HTML}{F0F2EE}
\newcommand{\nfText}{\fontfamily{phv}\fontsize{8.3}{10}\selectfont}
\newcommand{\nfSmall}{\fontfamily{phv}\fontsize{7.5}{9}\selectfont}
\newcommand{\nfTitle}{\fontfamily{phv}\bfseries\fontsize{10}{12}\selectfont}
\newcommand{\nfQuote}{\fontfamily{phv}\fontsize{9.5}{11.5}\selectfont}
\tikzset{
 nf/.style={line cap=round,line join=round,text=nfInk,font=\nfText},
 nf title/.style={font=\nfTitle,anchor=base west,inner sep=0pt},
 nf label/.style={font=\nfText,inner sep=1.5pt,align=left},
 nf small/.style={font=\nfSmall,text=nfGrey,inner sep=1.2pt,align=left},
 nf quote/.style={font=\nfQuote,inner sep=0pt,align=center},
 nf rule/.style={draw=nfLine,line width=.45pt},
 nf arrow/.style={draw=nfInk,line width=.8pt,-{Stealth[length=3.8pt,width=3.3pt]}},
 nf motion/.style={draw=nfAmber,line width=1.7pt,-{Stealth[length=5pt,width=4.6pt]}},
 nf lead/.style={draw=nfGrey,line width=.5pt},
 nf world/.style={x={(.80cm,-.35cm)},y={(0cm,.94cm)},z={(.80cm,.35cm)}},
 nf hand/.style={draw=nfInk,fill=white,line width=.85pt},
 nf tag/.style={font=\nfSmall,inner xsep=4pt,inner ysep=3pt,fill=nfBg,align=center},
}

\newcommand{\nfGateway}[2]{%
 \begin{scope}[nf world]
  \fill[nfPlane] (-.95,-.06,.6) -- (2.75,-.06,.6) -- (2.75,-.06,-1.85) -- (-.95,-.06,-1.85) -- cycle;
  \fill[nfStone!50!nfBg] (0,0,0) -- (1.8,0,0) -- (1.8,2.30,0) -- (0,2.30,0) -- cycle;
  \draw[fill=nfStone,draw=nfLine,line width=.55pt]
   (-.70,0,0) -- (0,0,0) -- (0,2.30,0) -- (1.8,2.30,0) -- (1.8,0,0) -- (2.5,0,0)
   -- (2.5,2.85,0) -- (2.1,2.85,0) -- (2.1,2.63,0) -- (1.7,2.63,0)
   -- (1.7,2.85,0) -- (1.3,2.85,0) -- (1.3,2.63,0) -- (.9,2.63,0)
   -- (.9,2.85,0) -- (.5,2.85,0) -- (.5,2.63,0) -- (.1,2.63,0)
   -- (.1,2.85,0) -- (-.3,2.85,0) -- (-.3,2.63,0) -- (-.7,2.63,0) -- cycle;
  \draw[fill=nfLine!55!nfBg,draw=nfLine,line width=.45pt] (2.5,0,0) -- (2.5,0,.28) -- (2.5,2.85,.28) -- (2.5,2.85,0) -- cycle;
  \draw[fill=white,draw=nfLine,line width=.45pt] (0,0,0) -- (0,0,.24) -- (0,2.3,.24) -- (0,2.3,0) -- cycle;
  \foreach \h in {.55,1.1,1.65,2.2}{
   \draw[nf rule] (-.7,\h,0) -- (0,\h,0) (1.8,\h,0) -- (2.5,\h,0);}
  \draw[nf rule] (0,0,0) -- (1.8,0,0);
  \ifnum#2>0
   \pgfmathsetmacro{\dx}{1.8*cos(#1)}\pgfmathsetmacro{\dz}{-1.8*sin(#1)}
   \ifnum#2=1
    \draw[draw=nfBlue,fill=nfBlue!8!nfBg,dash pattern=on 3pt off 2pt,line width=.8pt]
     (0,0,0) -- (\dx,0,\dz) -- (\dx,2.24,\dz) -- (0,2.24,0) -- cycle;
   \else
    \draw[draw=nfInk,fill=nfTeal!22!nfBg,line width=.9pt]
     (0,0,0) -- (\dx,0,\dz) -- (\dx,2.24,\dz) -- (0,2.24,0) -- cycle;
    \foreach \h in {.22,1.12,2.03}{\draw[draw=nfTeal!65,line width=.45pt]
      ({.12*\dx},\h,{.12*\dz}) -- ({.88*\dx},\h,{.88*\dz});}
    \draw[draw=nfTeal!65,line width=.45pt] ({.12*\dx},.22,{.12*\dz}) -- ({.12*\dx},2.03,{.12*\dz})
     ({.88*\dx},.22,{.88*\dz}) -- ({.88*\dx},2.03,{.88*\dz});
    \fill[nfInk] ({.83*\dx},1.06,{.83*\dz}) circle[radius=.038];
   \fi
   \draw[draw=nfInk,line width=1.25pt] (0,.06,0) -- (0,2.20,0);
  \fi
 \end{scope}}

\newcommand{\nfSweep}[2]{%
 \begin{scope}[nf world]
  \draw[nf motion] plot[variable=\a,domain=2:#1,samples=32,smooth]
   ({1.8*cos(\a)},#2,{-1.8*sin(\a)});
 \end{scope}}

\newcommand{\nfHand}{%
 \path[nf hand]
  (.22,0) -- (.19,.35) .. controls (.02,.56) and (-.03,.79) .. (-.12,.98)
  .. controls (-.23,1.20) and (-.06,1.30) .. (.08,1.10) -- (.25,.90)
  -- (.20,1.78) .. controls (.19,1.97) and (.39,1.98) .. (.40,1.80) -- (.45,1.22)
  -- (.45,2.01) .. controls (.45,2.19) and (.66,2.19) .. (.66,2.01) -- (.68,1.24)
  -- (.72,1.89) .. controls (.74,2.06) and (.93,2.04) .. (.92,1.87) -- (.90,1.16)
  -- (1.00,1.65) .. controls (1.04,1.79) and (1.22,1.73) .. (1.18,1.56)
  -- (1.07,.86) .. controls (1.05,.62) and (.91,.47) .. (.85,.32) -- (.84,0) -- cycle;
 \draw[draw=nfGrey,line width=.5pt] (.30,.76) .. controls (.45,.87) and (.75,.86) .. (.91,.77)
  (.35,.56) .. controls (.50,.68) and (.74,.64) .. (.85,.57);
 \draw[draw=nfInk,line width=.65pt] (.21,.15) -- (.84,.15);}

\newcommand{\nfTower}[2]{%
 \fill[nfBg] (-.18,-.06) rectangle (1.86,.04);
 \draw[fill=nfStone,draw=nfLine,line width=.5pt] (.94,0) rectangle (1.65,.92);
 \draw[fill=#1!17!nfBg,draw=#1,line width=.85pt] (.16,0) rectangle (.95,#2);
 \draw[fill=nfInk!13!nfBg,draw=nfGrey,line width=.6pt] (.08,#2) -- (.56,{#2+.48}) -- (1.03,#2) -- cycle;
 \draw[fill=white,draw=#1,line width=.55pt] (.45,{#2-.46}) rectangle (.65,{#2-.17});
 \draw[draw=#1!45,line width=.4pt] (.16,.38) -- (.95,.38) (.16,.77) -- (.95,.77);}

\ifdefined\ReviewSubmission
\setcopyright{none}
\else
\setcopyright{acmlicensed}
\fi
\copyrightyear{2027}
\acmYear{2027}

\acmDOI{} 

\acmConference[CHI '27]{CHI Conference on Human Factors in Computing Systems}{May 10--14, 2027}{Pittsburgh, PA, USA}

\acmISBN{}

\usepackage{xspace}

\usepackage{listings}

\lstdefinestyle{mdprompt}{
  basicstyle=\ttfamily\fontsize{6.6}{7.9}\selectfont\color{mdInk},
  backgroundcolor=\color{mdSofter},
  frame=single, rulecolor=\color{mdRule}, framerule=0.4pt, framesep=5pt,
  breaklines=true, breakindent=0pt,
  postbreak=\mbox{\textcolor{mdGrey}{$\hookrightarrow$}\,},
  columns=fullflexible, keepspaces=true, upquote=true,
  showstringspaces=false,
  aboveskip=9pt, belowskip=9pt,
  captionpos=t, abovecaptionskip=4pt, belowcaptionskip=4pt,
}

\newcommand{\redacted}[1]{\textsc{[redacted]}}

\begin{document}

\title{``Just Like This'': Manner Deixis at the Graphics Interface}

\begin{abstract}
People communicate how things should move by combining words with demonstrations: ``open it like this.'' We present an interaction technique that brings this expressive resource to conversational 3D authoring. Building on ``Put-That-There,'' our system combines speech, pointing, and spatiotemporal demonstrations to specify editable behavior. A hand movement supplies evidence for a mechanism's axis, pivot, range, and pace; an animated preview makes the interpretation inspectable. Users refine behavior through further words or demonstrations, and the system can request a demonstration to clarify intent. In a twelve-participant study comparing three input configurations, our system achieved 89\% participant-declared completion versus 47\% with speech alone, had the highest observed match rates on six categorical accuracy measures and tied on two, and was preferred by nine participants. This work makes demonstration part of an ongoing authoring conversation: behavior can be shown, inspected, and revised.
\end{abstract}



\author{Hamza El Alaoui} 
\orcid{0000-0001-9826-583X} 
\authornote{This work was conducted while the author was at Sony CSL.} 
\affiliation{%
  \institution{Sony CSL} 
  \city{Kyoto} 
  \country{Japan} 
} 
\additionalaffiliation{%
  \department{the School of Computer Science}
  \institution{Carnegie Mellon University} 
  \city{Pittsburgh} 
  \state{Pennsylvania} 
  \country{USA} 
}
\email{helalaou@cs.cmu.edu}

\author{Jeffrey Bigham} 
\orcid{0000-0002-2072-0625}

\affiliation{%
  \institution{Carnegie Mellon University} 
  \city{Pittsburgh, PA} 
  \country{USA}
} 
\email{jbigham@cs.cmu.edu}

\additionalaffiliation{%
  \department{the Artificial Intelligence/Machine Learning Group}
  \institution{Apple Inc} 
  \city{Pittsburgh} 
  \state{Pennsylvania} 
  \country{USA} 
}

\author{Jun Rekimoto} 
\orcid{0000-0002-3629-2514} 
\affiliation{%
  \institution{Sony CSL} 
  \city{Kyoto} 
  \country{Japan}
} 
\additionalaffiliation{%
  \department{the School of Information Studies}
  \institution{The University of Tokyo} 
  \city{Tokyo} 
  \country{Japan} 
} 
\email{rekimoto@acm.org}

\begin{teaserfigure}
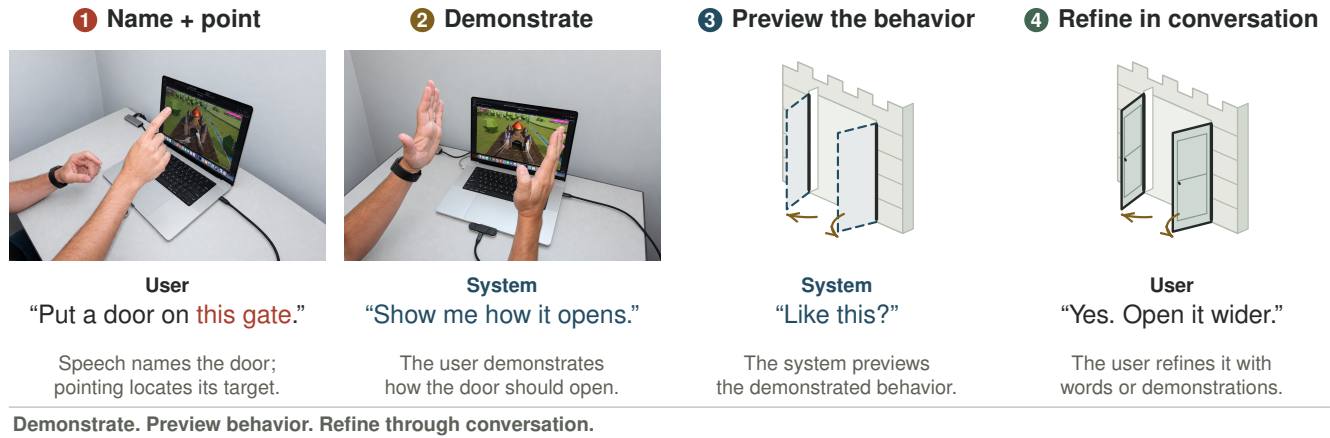

  \centering
  \ifdefined\ReviewSubmission
    \resizebox{\linewidth}{!}{\input{teaser}}
  \else
    \input{teaser}
  \fi
  \caption{Conversational behavior authoring: the user names and points to a target,
demonstrates how the door should open, approves its preview, and asks for a wider opening.
Photographs are retouched; diagrams are illustrative.}
  \Description{Four panels alternate user and system turns: a pointing hand at a laptop, with
the tracker partly obscured by the arm; two hands demonstrating above the
tracker; a dashed double-door preview; and two solid approved door leaves,
hinged on opposite sides with a gap between them. The system
requests the demonstration and confirmation; the user approves and asks for
a wider opening.}
  \label{fig:teaser}
\end{teaserfigure}

\maketitle

\section{Introduction}

Computers and interactive devices accept input through keyboards, pointers,
touch, gaze, and speech. Human communication draws on another expressive
resource: the movements people make while speaking. A hand can trace a
trajectory, enact an opening, or convey the rhythm of a wingbeat. Such
movements can carry spatial and temporal information that accompanying
words leave unspecified~\cite{mcneill1992hand,kita2003gesture,beattie1999iconic}.
Such movements benefit comprehension especially when depicting motor
content~\cite{hostetter2011gestures}. The opportunity for interaction design
is to make this expressive resource available when people communicate
behavior to a computer.

Bolt's \emph{Put-That-There} showed how speech and pointing could jointly
specify an action and its referents at the graphics interface~\cite{bolt1980put}.
Subsequent multimodal systems developed this combination of language and
spatial reference~\cite{cohen1997quickset,oviatt1999myths}. Yet a reference
alone does not specify a behavior: ``open that door'' identifies an action
and an object, while ``open it like this'' can use a hand movement to convey
how it should unfold. In linguistics, \emph{manner deixis} describes
expressions such as ``like this'' that direct attention to how an action is
performed; an accompanying demonstration can supply that
meaning~\cite{konig2018manner,guerin2015verbs}.

Demonstration-based animation and co-speech interfaces already use hand
movement to create or revise digital content~\cite{arora2019magicalhands,hu2025gesprompt,duan2026justshape}.
Our question is how demonstration can become part of an ongoing authoring
conversation: can speech, reference, and movement form one request whose
result remains inspectable and editable, and whose ambiguity can invite
another demonstration? This matters for language-driven authoring tools,
where users must express the behavior they want in a form the system can
interpret~\cite{delatorre2024llmr,yang2024holodeck}. Users of LLM animation
tools adapt prompts and goals to what they can verbalize~\cite{tseng2024keyframer},
while combining language with direct selection can reduce editing
effort~\cite{masson2024directgpt}. We extend this relationship to articulated
behavior: movement supplies a specification that later words and demonstrations
can revise.

We present an interaction technique for authoring 3D behavior through
speech, pointing, and demonstrated motion (Figure~\ref{fig:channels}).
We use \emph{demonstration} for a hand movement offered as an example of
the intended behavior.
Speech supplies intent and symbolic values; pointing grounds the target;
two-handed span supplies extent; and movement supplies evidence about
shape and pace. For example, a user says ``put a door on the gate,
opening like this'' and traces an arc. Our system summarizes that motion
as evidence for an articulated mechanism: a candidate hinge axis, pivot,
range, and pace. The resulting joint remains editable through another
sentence or demonstration. Figure~\ref{fig:teaser} follows the interaction
from request to preview and refinement.

\begin{figure}[t]
  \centering
\begin{tikzpicture}[nf]
\useasboundingbox (0,0) rectangle (8.4,7.15);
\tikzset{
 nf motion/.style={draw=nfAmber,line width=.9pt,
  -{Straight Barb[length=3.7pt,width=3.5pt]}},
 channel lead/.style={line width=.5pt,rounded corners=2pt}
}
\node[nf quote,font=\fontfamily{phv}\fontsize{13}{15}\selectfont,anchor=north] at (4.2,7.05)
 {``Put \textcolor{nfBlue}{a door} on \textcolor{nfRed}{this gate},\\opening \textcolor{nfAmber}{like this}.''};
\draw[nf rule] (0,5.70) -- (8.4,5.70);

\begin{scope}[shift={(4.85,2.75)},scale=.86]
 \colorlet{nfPlane}{white}
 \nfGateway{75}{2}
 \begin{scope}[nf world]
  \draw[draw=nfRed,line width=.85pt]
   (0,0,0) -- (0,2.30,0) -- (1.8,2.30,0) -- (1.8,0,0);
 \end{scope}
 \nfSweep{75}{-.12}
\end{scope}

\node[nf title,text=nfBlue] at (.15,4.78) {Speech};
\node[nf small,anchor=base west] at (.15,4.39) {What to create};
\draw[channel lead,draw=nfBlue] (2.05,4.46) -- (3.20,4.46) -- (4.12,3.94);
\fill[nfBlue] (4.12,3.94) circle[radius=.027];

\node[nf title,text=nfRed] at (6.80,4.78) {Pointing};
\node[nf small,anchor=base west] at (6.80,4.39) {Which target};
\draw[channel lead,draw=nfRed] (6.60,4.46) -- (6.30,4.46) -- (6.09,4.07);
\fill[nfRed] (6.09,4.07) circle[radius=.027];

\begin{scope}[shift={(3.30,1.12)},scale=.43,rotate=-22]
 \nfHand
\end{scope}
\node[nf title,text=nfAmber] at (.15,1.90) {Demonstration};
\node[nf small,anchor=base west] at (.15,1.51) {How it moves};
\draw[channel lead,draw=nfAmber] (2.85,1.58) -- (3.39,1.58);

\draw[nf rule] (0,.60) -- (8.4,.60);
\node[nf small,anchor=base west] at (0,.24) {\textbf{One request.} Words, pointing, and demonstration supply its meaning.};
\end{tikzpicture}
  \caption{One request combines object type (speech), target location (pointing; place
deixis), and intended behavior (demonstration; manner deixis).}
  \Description{Around a gateway, blue speech identifies the door, red pointing identifies the
target gate, and an amber hand and arc specify its opening movement.}
  \label{fig:channels}
\end{figure}

Grounding makes the composite request inspectable~\cite{clark1991grounding}.
Reference highlights show the selected part, a ribbon echoes captured
motion, and a moving preview shows the inferred behavior. A confidence
gate turns uncertain commands into questions~\cite{horvitz1999principles},
including ``show me how it should open,'' answered by a watched hand
movement. Prior systems request demonstrations for GUI concepts and robot
skills~\cite{li2019pumice,cakmak2012questions,myers1987peridot}; here that
exchange supports continuous 3D motion inside generative authoring.
Articulated-model fitting is established~\cite{sturm2011kinematic}; our
contribution makes it conversational.

We implement the technique on a desktop with a commodity hand tracker.
A castle courtyard supports articulated mechanisms and part-level edits;
a lighting atelier supports placement and hand-drawn trajectories. A
speech-anchored window aligns reference and movement with the utterance,
and a reasoning model returns structured commands for live execution.
Only the castle is evaluated in the comparative study.

In a within-subjects study (N=12), we compare speech-only, speech-and-pointing,
and input with all channels on behavior-authoring tasks whose targets are played in
the scene as translucent ghosts rather than described in words, measuring
categorical target matches and continuous parameter errors alongside recorded
trial duration, words, channel use, and workload.
Participants finished 89\% of trials with our system against 47\% with
speech alone ($p_{\mathrm{Holm}}=.006$), and nine of twelve selected it in
the final forced choice. This is a comparison of complete input
configurations: our system adds both span and motion to pointing.
Pointing appeared in 79--81\% of task turns in the two hand conditions;
span and motion appeared in 32\% and 21\% of turns with our system. Among
clarifications from our system with a subsequent turn in the same trial, 93\%
of those turns carried pointing, span, or motion evidence. These observations characterize how
participants used the available channels. Adjusted tests did not establish differences in accuracy, trial duration,
or word count.

This paper contributes:
\begin{itemize}
    \item \textbf{Conversational authoring of demonstrated mechanisms.}
    We operationalize manner deixis~\cite{konig2018manner,guerin2015verbs}
    by combining speech and pointing with motion evidence for an editable
    axis, pivot, range, and period. The contribution is the authoring
    integration, building on modeling through hand input, demonstration-based
    animation, and articulated-model fitting.

    \item \textbf{A grounding loop that accepts demonstration as an answer.}
    Captured-path feedback and inferred-motion previews expose different
    stages of interpretation. A confidence gate and \emph{watch mode}
    connect clarification to a new hand demonstration within the same
    scene-editing conversation. Their effectiveness as separate design
    components remains to be tested.

    \item \textbf{A bounded evaluation of cumulative input configurations.}
    A within-subjects study (N=12) finds higher participant-declared completion
    with our system than with speech alone, alongside physical costs
    of hand input. Channel records and participant accounts describe use;
    accuracy, timing, verbal effort, and preference are exploratory outcomes.
\end{itemize}

\section{Related work}\label{sec:rw}

Our work connects multimodal reference, demonstration-based animation,
generative authoring, and clarification. The distinction concerns how
these capabilities form an ongoing conversation about editable behavior.

\subsection{Speech, reference, and manner deixis}\label{sec:rw:deixis}

Bolt's \emph{Put-That-There} combined speech and pointing to resolve an
action and its spatial referents~\cite{bolt1980put}. QuickSet developed
semantic fusion of speech and pen input under temporal and semantic
constraints~\cite{cohen1997quickset}. Oviatt showed that input modalities
can supply complementary information and disambiguate one
another~\cite{oviatt1999myths,oviatt1999mutual}. Complementarity also
underlies Air+Touch, where touch selects targets and segments in-air
finger movements that specify additional actions~\cite{chen2014airtouch}.
More recent systems such as GazePointAR use gaze and pointing to ground
spoken references for language-model-based assistants~\cite{lee2024gazepointar}.
Together, these precedents motivate combining precise reference with
expressive input.

Hand movement contributes more than reference. McNeill distinguishes pointing
from iconic gestures that depict actions and spatial
relationships~\cite{mcneill1992hand}; Kendon treats visible action as part
of the utterance~\cite{kendon2004gesture}. Speakers can distribute path,
manner, size, and relative position across speech and
hand movement~\cite{kita2003gesture,beattie1999iconic}. The benefit to
comprehension is especially pronounced for motor content and information
that complements accompanying words~\cite{hostetter2011gestures}. This
motivates treating movement as information to interpret alongside speech.

\emph{Manner deixis} names a linguistic relationship, rather than a
particular motion-recognition technique. Whereas place demonstratives
such as ``there'' direct attention to a location, manner demonstratives
such as ``like this'' direct attention to how something is
performed~\cite{konig2018manner,guerin2015verbs}. An accompanying bodily
demonstration can supply the content of that expression. Interactional
accounts describe the coordination of these expressions with enactment
and gaze~\cite{stukenbrock2017so,streeck2002so}. In our interface, ``open
it like this'' connects the spoken action and referent to a captured
movement. The design question is how that composite instruction becomes a
behavior that remains available for inspection and revision.

\subsection{Hand input for modeling and animation}\label{sec:rw:shape}

Hand input has long served to express form. Surface Drawing deposits
surface geometry through hand strokes~\cite{schkolne2001surface}; Teddy
interprets informal sketches into 3D models~\cite{igarashi1999teddy}; and
Data Miming uses descriptive hand movements to identify objects~\cite{holz2011miming}.
Mockup Builder and Shape-It-Up develop bimanual and mid-air modeling
workflows~\cite{dearaujo2013mockup,vinayak2013shapeitup}. Elicitation work
also finds spontaneous combinations of speech and hand movement in multi-object
AR settings~\cite{zhou2022eliciting}. These precedents show how informal
spatial input can specify structure or reference.

\label{sec:rw:demo}
Animation systems use performance to convey dynamics. Motion Doodles
interprets sketches as parameterized character motion~\cite{thorne2004doodles};
spatial keyframing records a cursor performance~\cite{igarashi2005spatial};
and K-Sketch supports animation through informal
input~\cite{davis2008ksketch}. Draco and Kitty turn sketched examples into
continuing effects and interactive
behaviors~\cite{kazi2014draco,kazi2014kitty}. 3D Puppetry transfers the
motion of physical toys to virtual scenes~\cite{held2012puppetry}, while
a tangible skeleton controls articulated rigs~\cite{jacobson2014tangible}.
Creature Teacher turns bimanual manipulation of character parts into
looping animations~\cite{fender2015creature}. HandAvatar optimizes mappings
from hand joints to non-humanoid avatar joints for posing and
animation~\cite{jiang2023handavatar}.

MagicalHands is a particularly relevant precedent: it explores mid-air
hand input for animating dynamic phenomena in VR, including mappings
from direct manipulation to abstract demonstration~\cite{arora2019magicalhands}.
ARAnimator uses user-defined device movements for character
animation~\cite{ye2020aranimator}; AniCraft couples hand motion to physical
proxies~\cite{li2024anicraft}; and SketchPlay maps spatial sketches and
hand movements to scene structure and physical behavior~\cite{zhang2025sketchplay}.
These systems already establish demonstration as an animation input and
include editable representations. Our emphasis is the conversational use
of a demonstrated mechanism: speech and pointing supply context, movement
supplies geometric and temporal evidence, and later requests address the
resulting behavior.

Programming by demonstration provides the broader precedent for inferring
intent from examples~\cite{cypher1993watch}, including robot
learning~\cite{argall2009survey}. Estimating articulation from observed
motion is also established~\cite{sturm2011kinematic}. We use a lightweight
geometric summary within an authoring dialogue; we do not contribute a new
articulation-estimation algorithm.

\subsection{Language and hand input in generative authoring}\label{sec:rw:llm}

Language models support creative coding, visual design, and scene
generation in Spellburst, Inkspire, LLMR, and
Holodeck~\cite{angert2023spellburst,lin2025inkspire,delatorre2024llmr,yang2024holodeck}.
Their interfaces raise a common question: how can users specify and refine
intent beyond a verbal prompt? Keyframer describes iterative prompting
and direct editing in animation creation~\cite{tseng2024keyframer};
LogoMotion uses visually grounded program repair~\cite{liu2025logomotion};
and Narrative Motion Blocks combines language with direct
manipulation~\cite{bourgault2025blocks}. DirectGPT demonstrates the value
of selecting an edit target directly~\cite{masson2024directgpt}.
VRCopilot similarly combines speech, pointing, and drawn wireframes in
room-layout authoring~\cite{zhang2024vrcopilot}.

The closest systems already interpret hand input accompanying speech. GesPrompt
extracts spatial and temporal parameters, including rotation and paths,
from hand movements aligned with speech, and requests clarification for missing
or invalid parameters~\cite{hu2025gesprompt}. Its discussion also envisages
specifying an opening through demonstration. Thus neither motion demonstration nor asking a
clarifying question is unique to our work. JustShape combines hand input
with language for parametric CAD operations~\cite{duan2026justshape}.
GestuProp uses gripping hand poses and speech to specify VR
props~\cite{yao2026gestuprop}. HOICraft assists designers in configuring
part-level hand--object interaction mappings~\cite{lee2026hoicraft}.

Our focus is the articulated behavior being authored: a demonstrated
movement supplies evidence for a joint's axis, pivot, range, and period.
Those parameters remain addressable through subsequent speech or motion.
When more evidence is needed, the system can request a demonstration,
capture it without additional speech, and return an animated preview
within the same conversation. This integrated workflow extends the above
precedents; our study compares input configurations within it, not
performance against those published systems.

\subsection{Grounding and demonstration requests}\label{sec:rw:grounding}

Clark and Brennan describe grounding as obtaining evidence of understanding
sufficient for the current purpose~\cite{clark1991grounding}. Horvitz's
mixed-initiative principles connect uncertainty with decisions about when
to ask the user~\cite{horvitz1999principles}. Clarifying questions and
visible interpretations also support generative interfaces, including
text-to-image creation and 3D
editing~\cite{mehrabi2023resolving,hahn2025proactive,chen2026beyond}.

Requests to demonstrate have an established history. Peridot negotiates
inferences from demonstrated interface examples through confirming
questions~\cite{myers1986demonstration,myers1987peridot}. PUMICE asks users
to explain unfamiliar concepts and accepts GUI
demonstrations~\cite{li2019pumice}. Robot learners can actively request
new demonstrations~\cite{cakmak2012questions}. Our setting makes a
continuous hand movement an answer about 3D behavior. A captured-path
ribbon shows the received movement; an animated preview shows the inferred
mechanism. These distinct representations support inspection at different
stages of interpretation.

\subsection{Contribution in relation to prior work}\label{sec:rw:positioning}

The contribution is an interaction technique that connects three
established capabilities: interpreting speech with spatial reference,
abstracting demonstrated motion into editable behavior, and requesting
further demonstrations during clarification. Their connection lets an
author introduce behavior by showing it and continue discussing the
resulting mechanism. The implementation demonstrates this workflow; the
study evaluates the complete input configurations and characterizes how
participants used their channels. Neither the linguistic concept nor the
individual sensing, fitting, and clarification components is claimed as new.

\section{Our system}
\label{sec:system}

Our system supports desktop authoring of 3D scenes and their behaviors
by speaking, pointing, and demonstrating. A user sits at an ordinary display with
a hand tracker on the desk in front of them; a scene stands in the space
above it. They talk to the system as they would to a collaborator, naming
what they want, pointing to the target, and demonstrating motion by hand
to complement speech, with the scene updating in response.

\subsection{Design principles}
\label{sec:system:principles}

Four principles organize the interaction. \textbf{Combine complementary
evidence:} speech specifies categories and explicit values, pointing
supplies reference, and span and motion supply spatial and temporal
evidence (Figure~\ref{fig:channels}). The prompt prioritizes explicit
spoken values when they conflict with magnitudes inferred from hand input.
\textbf{Expose editable behavior:} articulated demonstrations become
joint parameters that later words or demonstrations can revise.
\textbf{Make interpretation visible:} reference highlights, captured paths,
and motion previews expose what was selected, received, and inferred.
\textbf{Ask for missing evidence:} clarification can request words,
pointing, or a demonstration, drawing on grounding and mixed-initiative
principles~\cite{clark1991grounding,horvitz1999principles}. These are design
policies; their implementation does not guarantee correct interpretation.

\subsection{Example interaction}
\label{sec:system:walkthrough}

This illustrative walkthrough follows one behavior from request to
revision; recorded study exchanges appear in
Section~\ref{sec:results:exchanges}. A user points to a castle gateway and
says ``put a door here.'' A highlight makes the target visible, but the
request leaves the opening unspecified. The system can ask for a
demonstration. The user traces a slow arc; a ribbon shows the captured path,
and a translucent door previews the inferred opening. Approval commits
the mechanism. The user then says ``slower,'' requesting a timing change
to that same hinge. The hand has returned to rest, but its demonstrated
behavior remains available for conversation.


\subsection{Architecture}
\label{sec:system:architecture}

\begin{figure}[t]
  \centering
\begin{tikzpicture}[nf]
\useasboundingbox (0,0) rectangle (8.4,8.35);
\node[nf title] at (0,8.02) {(a)\quad Bring the channels together};
\foreach \x/\c/\t/\s in {0/nfBlue/Speech/Intent,2.95/nfRed/Pointing/Reference,5.90/nfAmber/Demonstration/{Shape + pace}}{
 \draw[draw=nfLine,line width=.45pt,fill=nfBg] (\x,6.65) rectangle (\x+2.5,7.58);
 \node[nf label,font=\nfText\bfseries,text=\c] at (\x+1.25,7.25) {\t};
 \node[nf small] at (\x+1.25,6.91) {\s};
}
\draw[nf lead] (1.25,6.65) -- (1.25,6.30) -- (7.15,6.30) -- (7.15,6.65);
\draw[nf arrow] (4.20,6.65) -- (4.20,5.84);
\draw[draw=nfLine,line width=.6pt,fill=nfBg] (1.05,3.88) rectangle (7.35,5.84);
\node[nf title] at (1.30,5.43) {(b)\quad Interpret behavior};
\node[nf small,anchor=base west] at (1.30,5.04) {Python service + model};
\node[nf label,font=\nfText\bfseries,text=nfTeal] at (4.20,4.58)
 {A hand arc becomes an editable hinge.};
\node[nf small] at (4.20,4.17) {Grounded in the scene and conversation.};
\draw[nf arrow,draw=nfTeal] (4.20,3.88) -- (4.20,3.38);

\node[nf title] at (0,3.04) {(c)\quad Preview + refine};
\node[nf small,anchor=base west] at (0,2.65) {Unity live scene};
\begin{scope}[shift={(1.82,1.00)},scale=.45]
 \colorlet{nfPlane}{white}
 \tikzset{nf motion/.style={draw=nfAmber,line width=.7pt,
  -{Straight Barb[length=2.8pt,width=2.6pt]}}}
 \nfGateway{75}{1}\nfSweep{75}{-.12}
\end{scope}
\node[nf small,font=\nfSmall\bfseries,text=nfBlue,anchor=base] at (5.62,2.81) {System};
\node[nf quote,text=nfBlue] at (5.62,2.51) {``Like this?''};
\node[nf small,font=\nfSmall\bfseries,text=nfInk,anchor=base] at (5.62,2.12) {User};
\node[nf quote,align=center] at (5.62,1.60) {``Yes, but match\\my hand's speed.''};
\node[nf small,font=\fontfamily{phv}\fontsize{6.5}{8}\selectfont,align=center] at (5.62,.85)
 {Preview the behavior.\\Refine it with words or demonstrations.};
\draw[nf arrow,draw=nfBlue,rounded corners=3pt]
 (7.48,1.60) -- (8.10,1.60) -- (8.10,4.85) -- (7.35,4.85);
\draw[nf rule] (0,.30) -- (8.4,.30);
\node[nf small,anchor=base west] at (0,0)
 {Demonstration becomes behavior you can keep editing.};
\end{tikzpicture}
  \caption{Conversational authoring loop: (a) speech, pointing, and demonstration specify a
request; (b) the service and model interpret it as an editable hinge; (c)
Unity previews the behavior for further refinement. Schematic door example.}
  \Description{Speech, pointing, and demonstration enter a Python service and model, which
interpret a hand arc as a hinge using scene and dialogue context. Unity
previews the door; a return arrow carries the user's approval and request to
match the speed of their hand into the next turn.}
  \label{fig:architecture}
\end{figure}
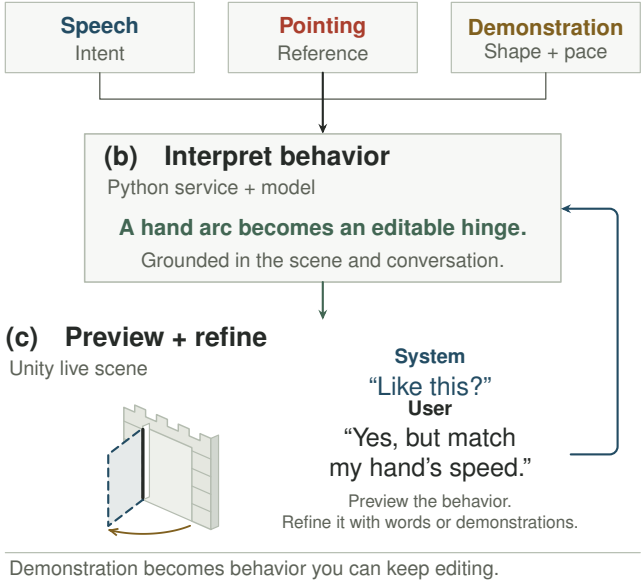

Our system comprises two processes joined by a local WebSocket
(Figure~\ref{fig:architecture}). A Unity front end owns sensing,
rendering, and execution. A Python service owns the microphone, speech
recognition, multimodal fusion, and dialogue state. The division follows the
second commitment: the front end evaluates and renders geometry locally; the service reasons
over language and structured scene and motion summaries, including fitted
coordinates, without directly editing mesh vertices.

The loop runs as follows. Unity streams deictic events and, on request,
clips of recorded hand motion. The service segments speech, transcribes it,
assembles a fusion prompt from the transcript plus summaries of the recent
deictic and motion evidence plus the current scene table, and calls a
reasoning model constrained to a strict JSON schema. The reply is validated,
gated on confidence, and forwarded to Unity, which highlights the resolved
referent, executes the command, and returns its execution status. The
service speaks and records that status, making execution feedback
available to subsequent turns.

\subsection{Sensing pointing and hand movement}
\label{sec:system:deictic}

The implementation supports near-surface fingertip contact and a
screen-space reference mapping; the study used the latter
(Section~\ref{sec:apparatus}). Contact feedback highlights the referenced
part. Per-vertex identifiers resolve mesh contact to named parts, while
props and room surfaces use stable identifiers on trigger volumes.

Unity emits 10\,Hz events for pointing, pinch transitions, and two-handed
span. Pointing includes the referent, world position, and dwell time;
span includes extent and axis. Interaction-mode flags help distinguish
reference from other movement. An eight-second ring buffer retains
stamped palm and fingertip positions, hand orientation, and grab and pinch
strengths, allowing a demonstration to be retrieved after speech ends.

The backend groups hovers into ordered episodes with rest times. This
preserves the sequence when a user refers to several objects within a turn.
Dwell provides contextual evidence to the reasoner rather
than deterministically resolving the target.

\subsection{Speech-anchored fusion}
\label{sec:system:fusion}

To avoid requiring a button press during two-handed input, the microphone
listens outside system-speech playback. Audio is captured at 16\,kHz in 32\,ms
blocks and segmented by a neural voice-activity detector (threshold 0.55,
600\,ms of trailing silence closes an utterance,
200\,ms of context retained on each side). Closed segments are
transcribed by a chunked speech-to-text call.

The fusion window is anchored to the \emph{utterance}, not to the moment
fusion runs. Voice-activity sample indices are mapped back to monotonic
capture time, and the window spans the speech interval padded by
1.5\,s before and 1.0\,s after. This retains reference evidence from the spoken request even when
transcription finishes later, and admits pointing immediately before or
after the corresponding words.

The assembled prompt contains the transcript, an ordered summary of pointing
episodes with dwell, an optional motion summary
(Section~\ref{sec:system:mechanism}), the scene table, and, when one is
open, the pending question or on-screen options. Structured replies specify
scene edits, object creation or updates, clarification questions, preview
selection, or speech without a scene change. The reply format constrains
structure; scene validation separately checks whether an action is executable.

%

\subsection{From demonstration to mechanism}
\label{sec:system:mechanism}

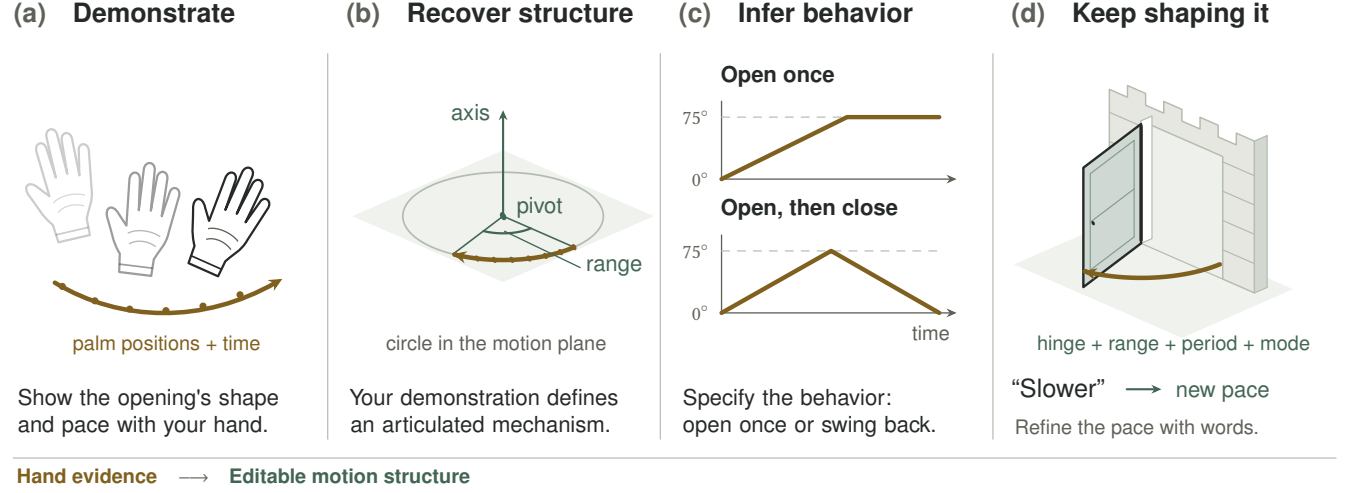
\begin{figure*}[t]
  \centering
  \ifdefined\ReviewSubmission
    \resizebox{\linewidth}{!}{
\begin{tikzpicture}[nf]
\useasboundingbox (0,0) rectangle (17.5,6.55);
\foreach \x/\n/\t in {0/a/{Demonstrate},4.40/b/{Recover structure},8.80/c/{Infer behavior},13.20/d/{Keep shaping it}}{
 \node[nf title] at (\x,6.17) {\textcolor{nfGrey}{(\n)}\quad\t};
}
\foreach \x in {4.16,8.56,12.96} \draw[nf rule] (\x,.65) -- (\x,5.75);
\begin{scope}[shift={(.45,3.22)},scale=.69,rotate=13,opacity=.24]\nfHand\end{scope}
\begin{scope}[shift={(1.25,2.88)},scale=.69,rotate=-7,opacity=.46]\nfHand\end{scope}
\begin{scope}[shift={(2.18,3.04)},scale=.69,rotate=-24]\nfHand\end{scope}
\draw[nf motion] (.55,2.72) .. controls (1.43,2.19) and (2.52,2.18) .. (3.56,2.77);
\foreach \x/\y in {.65/2.67,1.08/2.48,1.54/2.37,2.02/2.35,2.51/2.41,3.00/2.56}
 \fill[nfAmber] (\x,\y) circle[radius=.047];
\node[nf small,text=nfAmber,anchor=north] at (2.03,2.07) {palm positions + time};
\node[nf label,anchor=north west,text width=3.72cm] at (0,1.37) {Show the opening\textquotesingle s shape\\and pace with your hand.};
\begin{scope}[shift={(6.48,3.60)},scale=.82,nf world]
 \fill[nfPlane] (-1.52,0,-1.52) -- (1.52,0,-1.52) -- (1.52,0,1.52) -- (-1.52,0,1.52) -- cycle;
 \draw[draw=nfLine,line width=.65pt] plot[variable=\a,domain=0:360,samples=70,smooth]
  ({1.43*cos(\a)},0,{-1.43*sin(\a)});
 \foreach \a in {0,75}\draw[draw=nfTeal,line width=.6pt] (0,0,0) -- ({1.43*cos(\a)},0,{-1.43*sin(\a)});
 \draw[nf motion] plot[variable=\a,domain=0:75,samples=30,smooth]
  ({1.43*cos(\a)},0,{-1.43*sin(\a)});
 \foreach \a in {0,10,20,30,40,50,60,70}
  \fill[nfAmber] ({1.43*cos(\a)},0,{-1.43*sin(\a)}) circle[radius=.05];
 \draw[draw=nfTeal,line width=.85pt] plot[variable=\a,domain=0:75,samples=25,smooth]
  ({.56*cos(\a)},0,{-.56*sin(\a)});
 \draw[nf arrow,draw=nfTeal] (0,0,0) -- (0,1.82,0);
 \fill[nfTeal] (0,0,0) circle[radius=.065];
\end{scope}
\node[nf label,text=nfTeal,anchor=east] at (6.37,4.98) {axis};
\node[nf label,text=nfTeal,anchor=west] at (6.60,3.71) {pivot};
\draw[nf lead,draw=nfTeal] (6.50,3.40) -- (7.50,2.92);
\node[nf label,text=nfTeal,anchor=west] at (7.52,2.92) {range};
\node[nf small,anchor=north] at (6.39,2.07) {circle in the motion plane};
\node[nf label,anchor=north west,text width=3.72cm] at (4.4,1.37) {Your demonstration defines\\an articulated mechanism.};
\node[nf label,anchor=base west,font=\nfText\bfseries] at (9.30,5.38) {Open once};
\node[nf label,anchor=base west,font=\nfText\bfseries] at (9.30,3.61) {Open, then close};
\foreach \y in {4.09,2.32}{
 \draw[nf arrow,draw=nfGrey,line width=.5pt] (9.37,\y) -- (12.48,\y);
 \draw[nf lead] (9.37,\y) -- (9.37,\y+1.03);
 \draw[nf rule,dashed] (9.37,\y+.82) -- (12.28,\y+.82);
 \node[nf small,anchor=east] at (9.26,\y) {$0^\circ$};
 \node[nf small,anchor=east] at (9.26,\y+.82) {$75^\circ$};
}
\draw[draw=nfAmber,line width=1.65pt] (9.37,4.09) -- (11.03,4.91) -- (12.25,4.91);
\draw[draw=nfAmber,line width=1.65pt] (9.37,2.32) -- (10.82,3.14) -- (12.25,2.32);
\node[nf small,anchor=base east] at (12.44,1.96) {time};
\node[nf label,anchor=north west,text width=3.72cm] at (8.8,1.37) {Specify the behavior:\\open once or swing back.};
\begin{scope}[shift={(14.92,3.24)},scale=.75]
 \nfGateway{75}{2}\nfSweep{75}{.30}
\end{scope}
\node[nf small,text=nfTeal,anchor=north] at (15.35,2.07) {hinge + range + period + mode};
\node[nf quote,anchor=base west] at (13.2,1.21) {``Slower''};
\draw[nf arrow,draw=nfTeal] (14.73,1.31) -- (15.19,1.31);
\node[nf label,anchor=base west,text=nfTeal] at (15.32,1.21) {new pace};
\node[nf small,anchor=base west] at (13.2,.72) {Refine the pace with words.};
\draw[nf rule] (0,.40) -- (17.5,.40);
\node[nf small,anchor=base west] at (0,.07) {\textcolor{nfAmber}{\textbf{Hand evidence}}\quad $\longrightarrow$\quad \textcolor{nfTeal}{\textbf{Editable motion structure}}};

\end{tikzpicture}}
  \else
\begin{tikzpicture}[nf]
\useasboundingbox (0,0) rectangle (17.5,6.55);
\foreach \x/\n/\t in {0/a/{Demonstrate},4.40/b/{Recover structure},8.80/c/{Infer behavior},13.20/d/{Keep shaping it}}{
 \node[nf title] at (\x,6.17) {\textcolor{nfGrey}{(\n)}\quad\t};
}
\foreach \x in {4.16,8.56,12.96} \draw[nf rule] (\x,.65) -- (\x,5.75);
\begin{scope}[shift={(.45,3.22)},scale=.69,rotate=13,opacity=.24]\nfHand\end{scope}
\begin{scope}[shift={(1.25,2.88)},scale=.69,rotate=-7,opacity=.46]\nfHand\end{scope}
\begin{scope}[shift={(2.18,3.04)},scale=.69,rotate=-24]\nfHand\end{scope}
\draw[nf motion] (.55,2.72) .. controls (1.43,2.19) and (2.52,2.18) .. (3.56,2.77);
\foreach \x/\y in {.65/2.67,1.08/2.48,1.54/2.37,2.02/2.35,2.51/2.41,3.00/2.56}
 \fill[nfAmber] (\x,\y) circle[radius=.047];
\node[nf small,text=nfAmber,anchor=north] at (2.03,2.07) {palm positions + time};
\node[nf label,anchor=north west,text width=3.72cm] at (0,1.37) {Show the opening\textquotesingle s shape\\and pace with your hand.};
\begin{scope}[shift={(6.48,3.60)},scale=.82,nf world]
 \fill[nfPlane] (-1.52,0,-1.52) -- (1.52,0,-1.52) -- (1.52,0,1.52) -- (-1.52,0,1.52) -- cycle;
 \draw[draw=nfLine,line width=.65pt] plot[variable=\a,domain=0:360,samples=70,smooth]
  ({1.43*cos(\a)},0,{-1.43*sin(\a)});
 \foreach \a in {0,75}\draw[draw=nfTeal,line width=.6pt] (0,0,0) -- ({1.43*cos(\a)},0,{-1.43*sin(\a)});
 \draw[nf motion] plot[variable=\a,domain=0:75,samples=30,smooth]
  ({1.43*cos(\a)},0,{-1.43*sin(\a)});
 \foreach \a in {0,10,20,30,40,50,60,70}
  \fill[nfAmber] ({1.43*cos(\a)},0,{-1.43*sin(\a)}) circle[radius=.05];
 \draw[draw=nfTeal,line width=.85pt] plot[variable=\a,domain=0:75,samples=25,smooth]
  ({.56*cos(\a)},0,{-.56*sin(\a)});
 \draw[nf arrow,draw=nfTeal] (0,0,0) -- (0,1.82,0);
 \fill[nfTeal] (0,0,0) circle[radius=.065];
\end{scope}
\node[nf label,text=nfTeal,anchor=east] at (6.37,4.98) {axis};
\node[nf label,text=nfTeal,anchor=west] at (6.60,3.71) {pivot};
\draw[nf lead,draw=nfTeal] (6.50,3.40) -- (7.50,2.92);
\node[nf label,text=nfTeal,anchor=west] at (7.52,2.92) {range};
\node[nf small,anchor=north] at (6.39,2.07) {circle in the motion plane};
\node[nf label,anchor=north west,text width=3.72cm] at (4.4,1.37) {Your demonstration defines\\an articulated mechanism.};
\node[nf label,anchor=base west,font=\nfText\bfseries] at (9.30,5.38) {Open once};
\node[nf label,anchor=base west,font=\nfText\bfseries] at (9.30,3.61) {Open, then close};
\foreach \y in {4.09,2.32}{
 \draw[nf arrow,draw=nfGrey,line width=.5pt] (9.37,\y) -- (12.48,\y);
 \draw[nf lead] (9.37,\y) -- (9.37,\y+1.03);
 \draw[nf rule,dashed] (9.37,\y+.82) -- (12.28,\y+.82);
 \node[nf small,anchor=east] at (9.26,\y) {$0^\circ$};
 \node[nf small,anchor=east] at (9.26,\y+.82) {$75^\circ$};
}
\draw[draw=nfAmber,line width=1.65pt] (9.37,4.09) -- (11.03,4.91) -- (12.25,4.91);
\draw[draw=nfAmber,line width=1.65pt] (9.37,2.32) -- (10.82,3.14) -- (12.25,2.32);
\node[nf small,anchor=base east] at (12.44,1.96) {time};
\node[nf label,anchor=north west,text width=3.72cm] at (8.8,1.37) {Specify the behavior:\\open once or swing back.};
\begin{scope}[shift={(14.92,3.24)},scale=.75]
 \nfGateway{75}{2}\nfSweep{75}{.30}
\end{scope}
\node[nf small,text=nfTeal,anchor=north] at (15.35,2.07) {hinge + range + period + mode};
\node[nf quote,anchor=base west] at (13.2,1.21) {``Slower''};
\draw[nf arrow,draw=nfTeal] (14.73,1.31) -- (15.19,1.31);
\node[nf label,anchor=base west,text=nfTeal] at (15.32,1.21) {new pace};
\node[nf small,anchor=base west] at (13.2,.72) {Refine the pace with words.};
\draw[nf rule] (0,.40) -- (17.5,.40);
\node[nf small,anchor=base west] at (0,.07) {\textcolor{nfAmber}{\textbf{Hand evidence}}\quad $\longrightarrow$\quad \textcolor{nfTeal}{\textbf{Editable motion structure}}};

\end{tikzpicture}
  \fi
  \caption{Demonstration becomes editable motion: (a) opening shape and pace; (b) fitted
hinge axis, pivot, and range; (c) opening once or opening and closing; (d)
later pace refinement. Geometry and angle--time plots are illustrative.}
  \Description{Four panels show sampled hand poses; a fitted circle with axis, pivot, and
range; two illustrative angle-time curves with equal 75-degree ranges, one
holding and one returning; and a door retimed by a spoken correction.}
  \label{fig:mechanism}
\end{figure*}

A hand arc could specify a door's opening or a bird's route. Its role
depends on language and scene context; later edits also require an explicit
description of the inferred behavior.

When an utterance references a movement (``like this,'' ``open it this
way,'' or a bare flap of the hands), the recorded clip is summarized for the
model in three complementary forms. First, each palm's path is reduced to a
centered sixteen-point polyline in centimeters, so the model can read the
\emph{shape} itself rather than trusting our statistics. Second, per-axis
amplitude and frequency, posture (grab and pinch), and the two-hand gap and
its trend are reported. Amplitude is half the observed peak-to-peak range;
frequency is the number of crossings of the sample mean divided by twice
the full clip duration, a rough estimate for short or nonperiodic motion.
The gap trend flags mirrored-pair candidates such as
French doors. Third, and most consequentially, a fitted \emph{articulation}
block reports geometry (Figure~\ref{fig:mechanism}):

\begin{itemize}
\item An algebraic least-squares circle fit in a plane estimated from the
path's winding, whose
\textbf{normal} is the hinge-axis candidate and whose \textbf{center} is the
pivot, reported both in world coordinates and relative to the object.
\item \textbf{Net sweep} versus \textbf{total travel}. One complete opening
\emph{and} closing nets near zero while traveling twice the one-way range.
\item \textbf{Pace}, in degrees per second for arcs and centimeters per second
for straight paths, together with the clip duration.
\item \textbf{Straightness}, the chord-to-path-length ratio, which gates
whether an arc fit is attempted.
\end{itemize}

Concretely, let $\{(\mathbf{p}_i, t_i)\}_{i=1}^{n}$ be the sampled palm
positions in meters and their timestamps in seconds, $\bar{\mathbf{p}}$ their centroid, and
$\mathbf{q}_i = \mathbf{p}_i - \bar{\mathbf{p}}$. The plane of the movement
is estimated from the winding of the path about its own centroid. This is
not a least-squares plane fit. For nonzero winding, its normal is the
candidate hinge axis:

\begin{equation}
\mathbf{w}=\sum_{i=1}^{n-1}\mathbf{q}_i\times\mathbf{q}_{i+1},
\qquad
\hat{\mathbf{n}}=\frac{\mathbf{w}}{\lVert\mathbf{w}\rVert}.
\label{eq:normal}
\end{equation}

The implementation requires at least eight samples, path length
$L=\sum_{i=1}^{n-1}\lVert\mathbf{p}_{i+1}-\mathbf{p}_i\rVert\geq0.04\,\mathrm{m}$,
and $\lVert\mathbf{w}\rVert>10^{-8}\,\mathrm{m}^2$ before normalizing.
It also requires chord-to-path-length ratio
$\sigma=\lVert\mathbf{p}_n-\mathbf{p}_1\rVert/L<0.985$ to attempt an arc.
A uniformly sampled 45\textdegree{} circular arc has
$\sigma\approx0.9745$, so this gate admits short door swings. Near-straight
paths instead supply an endpoint direction and extent as slider evidence.
These are heuristics, not a complete arc/slider classifier: a straight
out-and-back path has $\sigma=0$, and a perfectly retraced curved path also
has zero net winding. In the latter case the arc fit is omitted; the
polyline and oscillation summaries remain available to the reasoner.

Within the estimated plane, let $\mathbf{u},\mathbf{v}$ be orthonormal basis
vectors, with $\mathbf{v}=\hat{\mathbf{n}}\times\mathbf{u}$, and let
$x_i=\mathbf{q}_i\cdot\mathbf{u}$ and $y_i=\mathbf{q}_i\cdot\mathbf{v}$.
The implementation obtains $\mathbf{u}$ by projecting the longest centered
sample into the plane and normalizing it, omitting the fit if that projection
is degenerate. We then solve the algebraic (Kasa) circle fit by linear
least squares:

\begin{equation}
\min_{a,b,c} \sum_{i=1}^{n}
\bigl( x_i^{2} + y_i^{2} - 2ax_i - 2by_i - c \bigr)^{2},
\qquad
r = \sqrt{\,c + a^{2} + b^{2}\,}.
\label{eq:kasa}
\end{equation}

The world-space pivot is
$\mathbf{c}_w=\bar{\mathbf{p}}+a\mathbf{u}+b\mathbf{v}$. For object
origin $\mathbf{o}$ and object-to-world rotation $R$ at unit scale, the
reported object-local pivot and axis are
$\mathbf{c}_o=R^{\mathsf T}(\mathbf{c}_w-\mathbf{o})$ and
$\mathbf{n}_o=R^{\mathsf T}\hat{\mathbf{n}}$.
A fit is accepted only when $r \in [0.02, 1.2]\,\mathrm{m}$ and the
mean absolute radial residual
$\frac{1}{n}\sum_{i=1}^{n}\left|\sqrt{(x_i-a)^2+(y_i-b)^2}-r\right|$
is below $0.25\,r$. This checks the projected circle; it does not bound
out-of-plane motion or guarantee a unique fit for degenerate data.
Let $\theta_i$ be the unwrapped angle $\operatorname{atan2}(y_i-b,x_i-a)$,
converted to degrees. With $D$ the full clip duration in seconds (floored at
0.001\,s in the implementation), the reported quantities are

\begin{equation}
\Delta\theta = \theta_n - \theta_1,
\qquad
T = \sum_{i=1}^{n-1} \lvert \theta_{i+1} - \theta_i \rvert,
\qquad
\omega = \frac{T}{D}.
\label{eq:sweep}
\end{equation}

A monotone one-way demonstration has $T=\lvert\Delta\theta\rvert$.
For one complete out-and-back excursion of one-way range $\psi$,
$\Delta\theta\approx0$ and $T\approx2\lvert\psi\rvert$. The reasoner uses
this distinction to propose opening once or swinging. The rule
$\lvert\psi\rvert\approx T/2$ applies only to a single excursion: $m$
repeated cycles traverse approximately $2m\lvert\psi\rvert$, so large
travel relative to net sweep alone cannot determine the range or cycle
count. Exact retracing can also suppress the arc summary, as noted above.
Here $\omega$ is a mean angular speed in degrees per second; it uses the
whole clip duration even if a hand is absent from some frames.

Pace is kept independent of geometry, which is what lets a later spoken
``slower'' retime a mechanism the hands built. For a one-sided swinging
hinge, the nominal full-cycle period is
$\tau=2\lvert\psi\rvert/\omega$; for a hinge swinging through both
$-\psi$ and $+\psi$, it is $\tau=4\lvert\psi\rvert/\omega$; and for a
spin it is $\tau=360^{\circ}/\omega$. Unity floors the requested speed at
1\textdegree/s and clamps each period to 0.4--12\,s. The one-time open
mode uses half its stored cycle period to open. Easing makes hinge speed
vary within the movement: these conversions match mean travel per unit
time, not constant instantaneous speed, and the period limits can prevent
an exact pace match. Angular-speed retiming leaves slider periods unchanged.

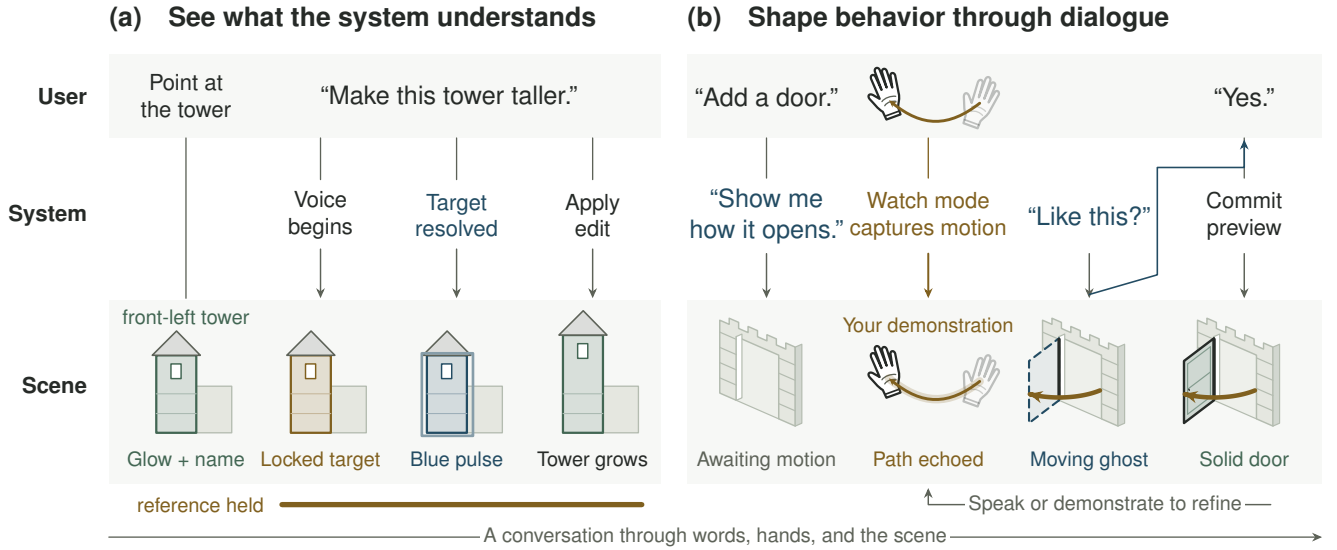
\begin{figure*}[t]
  \centering
  \ifdefined\ReviewSubmission
    \resizebox{\linewidth}{!}{
\begin{tikzpicture}[nf]
\colorlet{nfPlane}{nfBg}
\def\nfGroundingHandPath#1{%
 \begin{scope}[shift={(.40,.12)},scale=.32,rotate=-20,opacity=.28]\nfHand\end{scope}
 \begin{scope}[shift={(-.62,.12)},scale=.32,rotate=20]\nfHand\end{scope}
 \ifnum#1=1
  \draw[draw=nfAmber,line width=3.2pt,opacity=.20]
   (.68,.36) .. controls (.27,.04) and (-.08,.04) .. (-.52,.40);
 \fi
 \draw[nf motion,line width=1.1pt,-{Stealth[length=2.8pt,width=2.5pt]}]
  (.68,.36) .. controls (.27,.04) and (-.08,.04) .. (-.52,.40);
}
\useasboundingbox (0,0) rectangle (17.5,7.35);
\node[nf title] at (1.45,6.98) {(a)\quad See what the system understands};
\node[nf title] at (9.10,6.98) {(b)\quad Shape behavior through dialogue};
\fill[nfBg] (1.45,5.50) rectangle (8.75,6.60);
\fill[nfBg] (9.10,5.50) rectangle (17.5,6.60);
\fill[nfBg] (1.45,.92) rectangle (8.75,3.35);
\fill[nfBg] (9.10,.92) rectangle (17.5,3.35);
\foreach \y/\t in {6.05/User,4.48/System,2.23/Scene}
 \node[nf label,font=\nfText\bfseries,anchor=east] at (1.22,\y) {\t};
\node[nf label,align=center] at (2.47,6.06) {Point at\\the tower};
\node[nf quote] at (5.95,6.07) {``Make this tower taller.''};
\draw[nf lead] (2.47,5.50) -- (2.47,3.38);
\node[nf label,align=center] at (4.25,4.48) {Voice\\begins};
\node[nf label,align=center,text=nfBlue] at (6.05,4.48) {Target\\resolved};
\node[nf label,align=center] at (7.85,4.48) {Apply\\edit};
\foreach \x in {4.25,6.05,7.85}{
 \draw[nf lead] (\x,5.50) -- (\x,4.95);
 \draw[nf arrow,draw=nfGrey,line width=.55pt] (\x,4.03) -- (\x,3.38);}
\begin{scope}[shift={(1.97,1.60)},scale=.67]\nfTower{nfTeal}{1.53}\end{scope}
\begin{scope}[shift={(3.75,1.60)},scale=.67]\nfTower{nfAmber}{1.53}\end{scope}
\begin{scope}[shift={(5.55,1.60)},scale=.67]
 \nfTower{nfBlue}{1.53}
 \draw[draw=nfBlue!55,line width=1pt] (.06,-.06) rectangle (1.07,1.57);
\end{scope}
\begin{scope}[shift={(7.35,1.60)},scale=.67]\nfTower{nfTeal}{1.93}\end{scope}
\node[nf small,text=nfTeal] at (2.47,3.13) {front-left tower};
\foreach \x/\t/\c in {2.47/{Glow + name}/nfTeal,4.25/{Locked target}/nfAmber,6.05/{Blue pulse}/nfBlue,7.85/{Tower grows}/nfInk}
 \node[nf small,text=\c,align=center,anchor=north] at (\x,1.40) {\t};
\draw[draw=nfAmber,line width=2pt] (3.74,.65) -- (8.50,.65);
\node[nf small,anchor=east,text=nfAmber] at (3.55,.65) {reference held};
\node[nf quote] at (10.15,6.05) {``Add a door.''};
\begin{scope}[shift={(12.29,5.60)}]\nfGroundingHandPath{0}\end{scope}
\node[nf quote] at (16.47,6.05) {``Yes.''};
\node[nf quote,text=nfBlue] at (10.15,4.47) {``Show me\\how it opens.''};
\node[nf label,text=nfAmber,align=center] at (12.29,4.47) {Watch mode\\captures motion};
\node[nf quote,text=nfBlue] at (14.42,4.47) {``Like this?''};
\node[nf label,align=center] at (16.47,4.47) {Commit\\preview};
\foreach \x in {10.15,16.47}\draw[nf lead] (\x,5.50) -- (\x,4.97);
\draw[nf lead,draw=nfAmber] (12.29,5.50) -- (12.29,4.97);
\foreach \x in {10.15,14.42,16.47}\draw[nf arrow,draw=nfGrey,line width=.55pt] (\x,3.97) -- (\x,3.38);
\draw[nf arrow,draw=nfAmber,line width=.75pt] (12.29,3.97) -- (12.29,3.38);
\begin{scope}\clip (9.10,.92) rectangle (17.5,3.35);
\begin{scope}[shift={(9.75,2.03)},scale=.39]\nfGateway{0}{0}\end{scope}
\begin{scope}[shift={(12.29,1.92)}]\nfGroundingHandPath{1}\end{scope}
\node[nf small,text=nfAmber] at (12.29,3.03) {Your demonstration};
\begin{scope}[shift={(14.02,2.03)},scale=.39]\nfGateway{75}{1}\nfSweep{75}{1.05}\end{scope}
\begin{scope}[shift={(16.07,2.03)},scale=.39]\nfGateway{75}{2}\nfSweep{75}{1.05}\end{scope}\end{scope}
\foreach \x/\t/\c in {10.15/{Awaiting motion}/nfGrey,12.29/{Path echoed}/nfAmber,14.42/{Moving ghost}/nfBlue,16.47/{Solid door}/nfTeal}
 \node[nf small,text=\c,align=center,anchor=north] at (\x,1.40) {\t};
\draw[nf arrow,draw=nfBlue,line width=.6pt] (14.42,3.43) -- (15.32,3.69) -- (15.32,5.11) -- (16.47,5.11) -- (16.47,5.48);
\draw[nf arrow,draw=nfGrey,line width=.55pt] (16.82,.65) -- (12.29,.65) -- (12.29,.87);
\node[nf small,fill=white] at (14.63,.65) {Speak or demonstrate to refine};
\draw[nf arrow,draw=nfGrey,line width=.5pt] (1.45,.22) -- (17.5,.22);
\node[nf small,fill=white] at (9.47,.22) {A conversation through words, hands, and the scene};
\end{tikzpicture}}
  \else
\begin{tikzpicture}[nf]
\colorlet{nfPlane}{nfBg}
\def\nfGroundingHandPath#1{%
 \begin{scope}[shift={(.40,.12)},scale=.32,rotate=-20,opacity=.28]\nfHand\end{scope}
 \begin{scope}[shift={(-.62,.12)},scale=.32,rotate=20]\nfHand\end{scope}
 \ifnum#1=1
  \draw[draw=nfAmber,line width=3.2pt,opacity=.20]
   (.68,.36) .. controls (.27,.04) and (-.08,.04) .. (-.52,.40);
 \fi
 \draw[nf motion,line width=1.1pt,-{Stealth[length=2.8pt,width=2.5pt]}]
  (.68,.36) .. controls (.27,.04) and (-.08,.04) .. (-.52,.40);
}
\useasboundingbox (0,0) rectangle (17.5,7.35);
\node[nf title] at (1.45,6.98) {(a)\quad See what the system understands};
\node[nf title] at (9.10,6.98) {(b)\quad Shape behavior through dialogue};
\fill[nfBg] (1.45,5.50) rectangle (8.75,6.60);
\fill[nfBg] (9.10,5.50) rectangle (17.5,6.60);
\fill[nfBg] (1.45,.92) rectangle (8.75,3.35);
\fill[nfBg] (9.10,.92) rectangle (17.5,3.35);
\foreach \y/\t in {6.05/User,4.48/System,2.23/Scene}
 \node[nf label,font=\nfText\bfseries,anchor=east] at (1.22,\y) {\t};
\node[nf label,align=center] at (2.47,6.06) {Point at\\the tower};
\node[nf quote] at (5.95,6.07) {``Make this tower taller.''};
\draw[nf lead] (2.47,5.50) -- (2.47,3.38);
\node[nf label,align=center] at (4.25,4.48) {Voice\\begins};
\node[nf label,align=center,text=nfBlue] at (6.05,4.48) {Target\\resolved};
\node[nf label,align=center] at (7.85,4.48) {Apply\\edit};
\foreach \x in {4.25,6.05,7.85}{
 \draw[nf lead] (\x,5.50) -- (\x,4.95);
 \draw[nf arrow,draw=nfGrey,line width=.55pt] (\x,4.03) -- (\x,3.38);}
\begin{scope}[shift={(1.97,1.60)},scale=.67]\nfTower{nfTeal}{1.53}\end{scope}
\begin{scope}[shift={(3.75,1.60)},scale=.67]\nfTower{nfAmber}{1.53}\end{scope}
\begin{scope}[shift={(5.55,1.60)},scale=.67]
 \nfTower{nfBlue}{1.53}
 \draw[draw=nfBlue!55,line width=1pt] (.06,-.06) rectangle (1.07,1.57);
\end{scope}
\begin{scope}[shift={(7.35,1.60)},scale=.67]\nfTower{nfTeal}{1.93}\end{scope}
\node[nf small,text=nfTeal] at (2.47,3.13) {front-left tower};
\foreach \x/\t/\c in {2.47/{Glow + name}/nfTeal,4.25/{Locked target}/nfAmber,6.05/{Blue pulse}/nfBlue,7.85/{Tower grows}/nfInk}
 \node[nf small,text=\c,align=center,anchor=north] at (\x,1.40) {\t};
\draw[draw=nfAmber,line width=2pt] (3.74,.65) -- (8.50,.65);
\node[nf small,anchor=east,text=nfAmber] at (3.55,.65) {reference held};
\node[nf quote] at (10.15,6.05) {``Add a door.''};
\begin{scope}[shift={(12.29,5.60)}]\nfGroundingHandPath{0}\end{scope}
\node[nf quote] at (16.47,6.05) {``Yes.''};
\node[nf quote,text=nfBlue] at (10.15,4.47) {``Show me\\how it opens.''};
\node[nf label,text=nfAmber,align=center] at (12.29,4.47) {Watch mode\\captures motion};
\node[nf quote,text=nfBlue] at (14.42,4.47) {``Like this?''};
\node[nf label,align=center] at (16.47,4.47) {Commit\\preview};
\foreach \x in {10.15,16.47}\draw[nf lead] (\x,5.50) -- (\x,4.97);
\draw[nf lead,draw=nfAmber] (12.29,5.50) -- (12.29,4.97);
\foreach \x in {10.15,14.42,16.47}\draw[nf arrow,draw=nfGrey,line width=.55pt] (\x,3.97) -- (\x,3.38);
\draw[nf arrow,draw=nfAmber,line width=.75pt] (12.29,3.97) -- (12.29,3.38);
\begin{scope}\clip (9.10,.92) rectangle (17.5,3.35);
\begin{scope}[shift={(9.75,2.03)},scale=.39]\nfGateway{0}{0}\end{scope}
\begin{scope}[shift={(12.29,1.92)}]\nfGroundingHandPath{1}\end{scope}
\node[nf small,text=nfAmber] at (12.29,3.03) {Your demonstration};
\begin{scope}[shift={(14.02,2.03)},scale=.39]\nfGateway{75}{1}\nfSweep{75}{1.05}\end{scope}
\begin{scope}[shift={(16.07,2.03)},scale=.39]\nfGateway{75}{2}\nfSweep{75}{1.05}\end{scope}\end{scope}
\foreach \x/\t/\c in {10.15/{Awaiting motion}/nfGrey,12.29/{Path echoed}/nfAmber,14.42/{Moving ghost}/nfBlue,16.47/{Solid door}/nfTeal}
 \node[nf small,text=\c,align=center,anchor=north] at (\x,1.40) {\t};
\draw[nf arrow,draw=nfBlue,line width=.6pt] (14.42,3.43) -- (15.32,3.69) -- (15.32,5.11) -- (16.47,5.11) -- (16.47,5.48);
\draw[nf arrow,draw=nfGrey,line width=.55pt] (16.82,.65) -- (12.29,.65) -- (12.29,.87);
\node[nf small,fill=white] at (14.63,.65) {Speak or demonstrate to refine};
\draw[nf arrow,draw=nfGrey,line width=.5pt] (1.45,.22) -- (17.5,.22);
\node[nf small,fill=white] at (9.47,.22) {A conversation through words, hands, and the scene};
\end{tikzpicture}
  \fi
  \caption{Grounding through scene feedback: (a) contact, reference resolution, and
execution; (b) a requested demonstration, captured-path ribbon,
inferred-motion preview, and approval. Words or demonstrations continue the
conversation. Sequences are schematic.}
  \Description{User, system, and scene rows show two sequences. A tower highlights on
contact, locks gold at speech onset, pulses blue on resolution, then grows. A
door request prompts a demonstration; repeated hand poses align with its gold
path ribbon, followed by a dashed moving preview and a solid approved door. A
return arrow indicates refinement.}
  \label{fig:grounding}
\end{figure*}

We deliberately do \emph{not} classify the demonstration locally. The backend
reports numbers; the model decides what they mean in context, using the
scene, the words, and where the user was pointing.
Table~\ref{tab:taxonomy} gives the resulting mapping in full. This is a
consequence of the third commitment: the taxonomy is keyed to demonstrated
\emph{geometry} rather than to object names, so its reach is the reach of
the geometry itself. An arc about an edge is a hinge wherever it is drawn,
and the same fitted parameters drive a door, a lid, a wing, a jaw or a
book cover; a traced path is a trajectory whether the thing traveling is
a bird, a lamp or a cloud. The table names the readings the system makes,
not a catalog of objects it knows about.

Fitted parameters are written into a build specification: pieces grouped
into articulation groups, joints of type hinge, slider, or spin, each with a
pivot, axis, range, period, and mode (swing, open, or loop), which Unity
renders as animated pivot hierarchies. Because the mechanism is parametric,
later commands revise it: an angular pace retimes hinge and spin joints
from their authored ranges, while a requested period can retime all joint
types. A new demonstration can replace path and pace together.

\begin{table}[t]
\caption{Motion-to-behavior mappings, grouped by mechanism. Speech and scene context
guide interpretation; the examples do not prescribe fixed hand movements.}
\label{tab:taxonomy}
\small
\setlength{\tabcolsep}{4pt}
\renewcommand{\arraystretch}{1.03}
\begin{tabular}{@{}p{0.54\columnwidth}p{0.38\columnwidth}@{}}
\toprule
\textbf{What the hand did} & \textbf{What it becomes} \\
\midrule
\multicolumn{2}{@{}l}{\textsc{hinge} \textemdash{} rotation about an edge} \\[1pt]
\quad Horizontal arc about a vertical edge & Swing door \\
\quad Vertical arc, pivoting at its base & Drawbridge \\
\quad Vertical arc, pivoting at its top & Up-and-over door \\
\quad Two mirrored arcs, parting & French double doors \\
\quad Fast, small vertical oscillation & Wingbeat \\
\addlinespace[2pt]
\multicolumn{2}{@{}l}{\textsc{slider} \textemdash{} travel along a line} \\[1pt]
\quad Straight horizontal path & Sliding door \\
\quad Straight vertical path, upward & Portcullis \\
\quad Two straight paths, parting & Bi-parting sliders \\
\addlinespace[2pt]
\multicolumn{2}{@{}l}{\textsc{spin} \textemdash{} continuous rotation} \\[1pt]
\quad Full flat circle & Revolving door \\
\addlinespace[2pt]
\multicolumn{2}{@{}l}{\textsc{path and period} \textemdash{} the movement itself} \\[1pt]
\quad Traced open path & Trajectory, at that pace \\
\quad Slow side-to-side sway & Waving cloth, at that period \\
\bottomrule
\end{tabular}
\end{table}

\subsection{Grounding and clarification}
\label{sec:system:grounding}

Feedback makes reference, capture, and interpretation visible
(Figure~\ref{fig:grounding}). Contact highlights and names a part; speech
onset locks the reference; a pulse identifies the resolved target before
an edit. A ribbon shows the captured hand path, while an animated preview
shows the inferred behavior.

Commands below the model-reported confidence threshold become clarification
questions. For ambiguous reference, the interface highlights the model's
candidate and the most-dwelled alternative referent. Vague magnitudes produce
two or three previewed alternatives. The same gating applies to prop actions.

In \textbf{watch mode}, a clarification requests a demonstration and captures
its motion until stillness or timeout. The clip is fused with the original
request without requiring additional speech. New speech cancels the watch;
while it is active, movement cannot select a clarification option.

In \textbf{ghost preview}, a demonstrated object appears as a translucent,
animated preview. Approval commits it; rejection removes it and reopens the
request; correction revises it. Moving on is treated as acceptance.

The session retains the last six exchanges, including the user's words,
a compact description of each action, and its execution status. This
memory supports subsequent references such as ``make it slower.'' Bare
pronouns bind by recency among props, allowing later instructions to
continue working with the same object.


\subsection{Scene editing and behavior execution}
\label{sec:system:execution}

Unity executes semantic commands against named parts and a stable prop
ledger. Geometry edits select a part, a neighborhood around a touched
point, or the whole object; deformation is weighted within that selection
and committed as one undo step. A prop can be created, removed, restyled,
retimed, moved, or rebuilt while retaining a referenceable identifier.
Its live state returns to the scene summary for subsequent requests.
Articulated props use primitive pieces grouped under hinge, slider, or
spin joints. The reasoner supplies their composition and parameters;
Unity constructs and animates the resulting hierarchy.

The castle's doorway illustrates a maintained spatial relationship.
Its published origin and dimensions ground door creation. When a door's
size changes, the surrounding opening adapts within structural limits,
and updated dimensions return to the reasoner. This supports refinement
without requiring separate instructions for every adjoining part.

We instantiate the interface in a castle courtyard for mechanisms and
part-level editing, and a lighting atelier for placement, trajectories,
and pace. Each scene supplies its own prompt and scene description.
Only the castle is evaluated in the comparative study.

\subsection{Implementation and response timing}
\label{sec:system:implementation}

The front end is Unity 6000.0.79f1 with the Universal Render Pipeline and
the Ultraleap Unity plugin 7.3.0 over the Gemini 5.20 tracking service,
driving a desk-mounted Leap Motion Controller 2,
an optical tracker that reports hand and finger pose from infrared imagery
without anything worn or held; the
service is Python, using a neural voice-activity detector, a hosted
speech-to-text model, and a hosted reasoning model queried through a strict
JSON-schema interface, with spoken replies synthesized and gated against the
open microphone. Calibration preserves level and unit scale while estimating yaw and
translation, four free parameters. In a roughly five-second capture, the
user cups the virtual object with open hands: the line between the palms
defines screen-right and their midpoint defines object position. The
resulting pose is stored relative to the camera. Mouse input mirrors the
pointing channel as a reference fallback, following the condition's
available inputs (Section~\ref{sec:apparatus}).

Code and supplementary materials are described in
Appendix~\ref{app:prompts}.

The logged fusion interval begins immediately before the reasoner call and
ends at the turn record, after response handling. It excludes earlier
utterance closure and transcription. Across the 973 retained task turns,
its median was 4.01~s (IQR 2.79--5.78~s), its ninetieth percentile 8.20~s,
and its maximum 27.56~s. Condition medians were 3.62~s for speech alone,
3.85~s with pointing, and 4.33~s with our system. These summarize the
logged interval, not end-to-end response latency or an isolated estimate of
demonstration overhead.

Study interactions used live model calls without a Wizard-of-Oz operator.
Targets were predefined animations, and scene execution used the deterministic
operators described above. The code snapshot documents implementation;
the retained study tables document the reported observations.

\section{Study methods}
\label{sec:study}

We organize the evaluation around three questions:
\begin{description}
\item[RQ1: Success and effort.] How does adding pointing and
 demonstrated motion to speech affect task success and authoring effort
 when creating spatiotemporal behavior?
\item[RQ2: Complementary input channels.] How do speech, pointing, and
 demonstrated motion complement one another in expressing spatiotemporal behavior?
\item[RQ3: Clarification responses.] How do users respond to system-initiated
 clarification through pointing, span, and demonstrated motion?
\end{description}
We assess task success through participant-declared completion and target
accuracy, and authoring effort through recorded trial duration, word count,
and perceived workload. These are cumulative configurations: pointing is
added to speech, then span and motion are added together. The comparison
does not isolate a motion-only effect. RQ2 and RQ3 are descriptive: channel
logs and participant accounts characterize complementarity, while RQ3
examines which input channels appear in the next reply within the same
trial. It does not measure request resolution.
Preference provides supplementary evidence about the experience of using
each configuration.

\subsection{Participants}
\label{sec:participants}

We recruited 12 participants from the university community and the
surrounding area by email, printed flyers, and word of mouth. Participants included
students and working professionals from a range of fields. Age was collected in bands and ranged from 18 to 64, with
most participants (58\%) in the 18 to 24 band. Six identified as male and six as female, and two
were left-handed. 

All but one participant used 3D authoring tools a few times a year at most;
prior hand-tracking experience was generally limited. Half used voice
assistants at least monthly. Detailed backgrounds and self-reported
communication habits appear in Appendix~\ref{app:participants}.

Each session lasted 60 minutes and participants were compensated with a
\$20 gift card. The study was approved by \redacted{}. All
instructions, target descriptions, and system speech were in English, and
all participants reported being comfortable conversing in it.

\subsection{Apparatus and design}

\paragraph{Apparatus.}
\label{sec:apparatus}
Participants sat at a desk facing a single display of the castle scene.
A desk-mounted Ultraleap tracker observed their hands from below. An Apple
MacBook Pro (M3 Pro, 36\,GB) ran Unity, hand tracking, and voice-activity
detection, using its built-in microphone. Transcription, reasoning, and
speech synthesis used hosted models described in Appendix~\ref{app:params},
with models and settings held constant across conditions. Sessions took place in a quiet room with
the participant and experimenter present.

Fingertip reference was resolved in screen space to reduce the need to
judge contact depth on the flat display. Mouse input could provide the
same reference as a fallback; the pointing flag does not distinguish
these sources. Neither input directly manipulated geometry: participants
authored by requesting changes. The comparison concerns ways of specifying
behavior within a conversational interface.

Before practice, participants nominated a pointing hand, confirmed against
their reported handedness. Single-hand interactions followed that choice;
bimanual spans were symmetric. Calibration aligned the tracker and scene
from a relaxed two-handed pose. Participants checked pointing against
several castle parts and repeated calibration until satisfied. They could
recalibrate later in the session.

\paragraph{Input conditions.}
The study compares three input configurations of our system, with
\emph{available channels} as the within-subjects factor.
In \textbf{speech only (S)}, participants use speech; pointing, span, and
motion inputs are disabled. In \textbf{speech + pointing (P)}, pointing
also resolves reference, while span and motion remain unavailable.
With \textbf{our system (M)}, speech and pointing are joined by two-handed
span and demonstrated motion. Here, ``our system'' denotes the configuration with all channels enabled;
S and P are restricted input configurations of the same implementation.

All conditions use the same backend, feedback, and clarification machinery.
Channel availability is enforced before evidence reaches the reasoner.
Condition-specific instructions and output checks prevent the system from
requesting unavailable input, such as a demonstration in the speech-only
condition.

\paragraph{Counterbalancing.}
We counterbalanced condition order completely: all $3! = 6$ orders were
used, and we assigned them so that over each block of six participants every
condition appears equally often in every position and every ordered pair of
conditions appears equally often. Assignments were fixed before recruitment
and followed in order, so no ordering decision was left to the session
itself.

We counterbalanced the target sets on the same principle. Each participant
met all three sets, one per condition, and we crossed set against both
condition and position: over a block of six participants, every condition
meets every set exactly twice, and every set appears in every position
exactly twice. The design is therefore balanced on three crossings at once,
condition~$\times$~position, condition~$\times$~set, and
set~$\times$~position, with no participant seeing the same set twice. No
simple rotation satisfies all three simultaneously, so we constructed a
fixed schedule of six rows and applied it cyclically.
Balance is exact at every multiple of six.

\paragraph{Tasks and targets.}
Each participant completed three tasks in every condition, in the fixed
order \textbf{door}, \textbf{birds}, and \textbf{flag}, with a nominal
three-minute limit per task. We selected these tasks to span three
complementary forms of behavior specification: articulated mechanisms,
combined trajectory and local motion, and wind-like cloth motion. Together,
they test whether adding pointing and demonstration helps users convey
an intended behavior more effectively and efficiently than speech alone
or speech with pointing. Completion and target accuracy assess
 effectiveness; recorded duration, words spoken, and workload assess
 authoring effort.

The \textbf{door} task requires specifying how an object is constructed
and opens. Variants include sliding, paired French, and top-hinged doors,
which differ in their moving parts, articulation, and opening behavior.
The task therefore goes beyond naming a door style: users must communicate
how its parts should move.

The \textbf{birds} task combines movement through the scene with movement
of the wings. A flock can follow a circular route or fly back and forth
along a line, while its wings beat with different rates and amplitudes.
The targets combine differences in flock size, path, travel direction,
and wing motion, including slow, deep strokes and faster, shallower beats.
This task tests whether users can convey several coordinated aspects of
behavior: where the birds travel and how they move while following that path.

The \textbf{flag} task emphasizes the shape and rhythm of wind-like motion.
Targets depict a wave traveling along the cloth, the whole sheet flapping,
or the free edge curling, each with a different period. Users must convey
both where movement occurs on the flag and how it unfolds over time,
alongside its design and attachment location.

Three target sets each contained one variant of every task. Targets were
presented as replayable translucent animations in the scene, without a
verbal description, so participants could observe the intended behavior
without being supplied with language for requesting it. Condition order
and target sets were counterbalanced as described above; task order was
fixed. Variants were designed to exercise comparable kinds of specification
within each task, without assuming equal difficulty.

\subsection{Pilot study}
\label{sec:pilot}

Two pilot participants (P1 and P2), excluded from the main study and all
reported analyses, completed the protocol. The pilot established the
nominal three-minute task interval and session procedure. The conditions,
ghost targets, and questionnaires were fixed beforehand.

\subsection{Procedure}
\label{sec:procedure}

Each session lasted one hour with a participant and experimenter.
Participants signed consent covering the recordings and their right to
stop, then completed the background questionnaire summarized in
Section~\ref{sec:participants}.
%

\paragraph{Briefing.}
A fixed script introduced the three behavior tasks, the available channels,
the replayable ghost targets, and clarification. Participants were told to
work until satisfied or until three minutes had passed. Conditions were
named by their available inputs, without identifying a favored condition.

\paragraph{Practice.}
Before the first round, participants had three minutes of target-free
practice with all channels, regardless of condition order. They could
point, speak, and demonstrate to familiarize themselves with the feedback.
Practice data are excluded from analysis.

\paragraph{Task rounds.}
The three rounds then followed, one per condition, in the counterbalanced order. Each round comprised the three tasks in the fixed
order door, birds, flag, each against the target from the set assigned to
that round. At the start of a task the target ghost played once in the
scene; the participant could replay it from an on-screen button at any
time, and each replay was logged. The nominal three-minute task interval began when the
experimenter pressed Start. The experimenter recorded whether participants
finished, ran out of time, or gave up; the outcome is
recorded as one of these. The experimenter did not intervene during a task
except to restart one whose connection to the tracker had dropped; such a
task is recorded with the outcome \emph{disconnected}, restarted, and the
restart logged as a further attempt. Where a task has more than one attempt, analysis uses the
attempt retained in the trial table, including its recorded outcome.

\paragraph{Session controls.}
The experimenter's panel displayed the participant and planned next step.
Unavailable channels were removed before fusion, and the reasoner received
condition-specific instructions naming the available channels.

\paragraph{Questionnaires and closing.}
After each round, participants completed NASA-TLX, two CSI factors, eight
experience items, and two open questions. After all rounds, they ranked
the conditions, selected an overall preference, and answered three final
open questions. The instruments and questions are defined below.
%

\subsection{Measures}
Measures and results follow the same order: task outcomes, complementary
input channels, clarification responses, and workload and preference.

\subsubsection{Task outcomes (RQ1)}

\paragraph{Task outcome.}
The experimenter recorded one of three mutually exclusive outcomes:
\emph{finished} when participants declared satisfaction, \emph{ran out of
time} when stopped at the task limit, or \emph{gave up} when they stopped
early because they could not obtain the intended result. Completion is
participant-declared and distinct from target accuracy.

\paragraph{Time on task.}
Duration runs from the experimenter's Start action to logged trial closure,
regardless of outcome. Logged overruns beyond the nominal 180-second limit
are retained. Because abandonment can end a trial early, duration is
interpreted alongside completion, not as time to success. We also report
means by task.

\paragraph{Words spoken.}
Transcribed words per trial, counting only the participant. This
exploratory measure asks whether hand input is accompanied by less verbal
effort. Like duration, word count must be read alongside completion: early
abandonment can reduce the amount spoken.

\paragraph{Target accuracy.}
We compare retained final results with their targets, including unfinished
trials. Eight measures report whether the requested properties match:
door mechanism, moving leaves, and opening mode; flock path and direction;
and flag wave shape, design, and attachment. Six additional measures
quantify the absolute difference from the target in door axis angle and
period, flock count, wing-beat rate and amplitude, and flag wave period.
For example, wing-beat rate is compared in hertz and period in seconds;
smaller differences indicate closer agreement. Analyses use the retained
accuracy tables. The supplement reports the values, sample sizes, and tests
for all fourteen measures.

\subsubsection{Complementary input channels (RQ2)}

\paragraph{Logged channel evidence.}
Logs capture transcripts, system replies, scene commands, pointing episodes,
two-hand spans, and demonstrations. The turn table records whether pointing,
span, or motion evidence was available to fusion; channels may co-occur.
We report marginal presence rates, not causal attribution of parameters to
channels. Selected retained turn sequences illustrate how an authoring
request continues across modalities; they illustrate interaction patterns
rather than serving as a separately coded outcome. Pointing episodes group consecutive samples of the same referent,
ending when the referent changes or tracking is absent for one second.
Synchronized hand, screen, and tracker recordings support internal log
checks; they were not independently coded as outcome measures. Study
recordings do not appear in the paper; Figure~\ref{fig:teaser} uses author
photographs taken after the study.

\paragraph{Participant accounts.}
After each condition, two open questions ask about interaction challenges
and intentions the participant could not express. After all three,
participants explain their preferred condition, task-specific preferences,
and any intentions unexpressible across conditions. Responses are grouped
by channel use, task preferences, expressive scope, and interaction
preferences. Counts refer to main-study participants; pilot responses are
excluded.

%

\subsubsection{Clarification responses (RQ3)}
We report clarification frequency, requested response modality, and evidence
in the next logged turn within the same retained trial. A nonverbal reply
contains pointing, span, or motion evidence; exclusive counts prioritize
motion, then span, then pointing. These summaries characterize response
modalities rather than request resolution.


\subsubsection{Workload and preference}

\paragraph{Perceived workload.}
After each condition participants complete the NASA-TLX~\cite{hart1988nasatlx} on all six
subscales (mental demand, physical demand, temporal demand, performance,
effort and frustration), each a 0 to 100 slider in steps of five. We report
the raw TLX, the unweighted mean of the six, rather than running the
pairwise weighting procedure; the raw score correlates closely with the
weighted one and is the more common choice in recent
work~\cite{hart2006nasatlx}, and here the subscales are also reported
individually. Every slider runs from its light end to its heavy end,
performance included (0 perfect, 100 failed), so a higher number is a
heavier load on all six. We add one item outside the instrument, arm and
hand tiredness on a 0 to 10 scale, because two of the three conditions ask
for sustained hand use and the TLX's physical demand subscale is worded for
whole-body effort. It is reported separately and is not part of the raw
TLX.

\paragraph{Creativity support and experience.}
After each condition participants also complete two Creativity Support
Index~\cite{cherry2014csi} factors,
Expressiveness and Results Worth Effort, and eight custom seven-point
items covering expressiveness, the cost of putting an intention into words,
grounding, receipts, fidelity, workaround use, agency, and preference.
Each CSI factor averages two 1--10 items; the full weighted index was not
administered. These ratings are supplementary and descriptive.

\paragraph{Rankings and overall preference.}
After all conditions, participants rank them by ease of use, match to
intention, and willingness to use again, then choose one condition to keep.
Rankings express criterion-specific order, whereas the forced choice records
overall preference.

\subsection{Analysis}

The study was not preregistered. The RQs organize the reported analyses,
rather than constituting preregistered hypotheses. Completion is the primary
comparative outcome. RQ2 and RQ3 are addressed descriptively through channel-presence
rates, clarification replies, and participant accounts; we do not estimate
a channel-by-parameter attribution matrix. Duration, word count, workload,
accuracy, and preference analyses are exploratory.

For each comparative outcome, a participant contributes one value per
condition: their completion proportion or mean over retained tasks, or
their condition-level questionnaire rating. For our system's paired
completion gain over speech alone, we report a 95\% bias-corrected and accelerated
(BCa) bootstrap interval (100,000 participant resamples; no multiplicity
adjustment).
We use the tie-corrected
Friedman statistic with exact within-participant permutation probabilities,
and Kendall's $W=Q/[N(k-1)]$, with $k=3$. For significant omnibus outcomes,
pairwise two-sided Wilcoxon signed-rank tests use exact sign permutations,
omit zero differences, assign midranks to tied absolute differences, and
apply Holm correction across the three pairs. Pairwise effects are
rank-biserial correlations $r_{\mathrm{rb}}=(R_+-R_-)/(R_++R_-)$ over
nonzero differences: $r_{\mathrm{rb}}=1$ means all are positive,
excluding participants with equal scores.
We additionally adjust across the six TLX subscales
and, separately, all fourteen accuracy measures. Unadjusted exploratory
results are not treated as confirmatory evidence. Complete participant
triplets are used for each accuracy test; descriptive denominators and
paired sample sizes are retained in the supplementary summaries
(Appendix~\ref{app:prompts}).

Overall preference is tested against equal choice probabilities using an
exact multinomial test ordered by Pearson's $\chi^2$ statistic. Rankings
are analyzed separately with the same participant-level omnibus procedure.

Incomplete and abandoned trials count as failures for completion. Duration
and word counts retain the values in the trial table, and accuracy uses
available scored parameters without worst-case imputation. Thus the
handling of an incomplete trial depends on the outcome being measured.

Free-text responses are coded inductively by one author. The main-study
corpus contains 101 nonempty answers from twelve participants. We report
the grouping as an organization of participant accounts, not as a measured
construct; quotations make that interpretation inspectable. No inter-rater
reliability is claimed for this single-coder analysis.

\section{Results}
\label{sec:results}

Twelve participants attempted three tasks under each condition, giving 108
retained trials. Restarted sessions contribute the retained attempts in
the trial table. Turn analyses join to these same participant, attempt,
condition, task, and trial identifiers, yielding 973 task turns; turns
outside a retained task are excluded. Condition-level questionnaires use
one retained response per participant and condition.

RQ1 is addressed in Sections~\ref{sec:results:completion},
\ref{sec:results:effort}, \ref{sec:results:accuracy}, and~\ref{sec:results:workload}; RQ2 combines the
channel records in Section~\ref{sec:division} with participant accounts in
Section~\ref{sec:results:accounts}; and RQ3 is addressed in
Section~\ref{sec:grounding}. Preference provides supplementary context.

\begin{figure}[t]
\centering
\ifdefined\ReviewSubmission
\includegraphics[width=.68\linewidth]{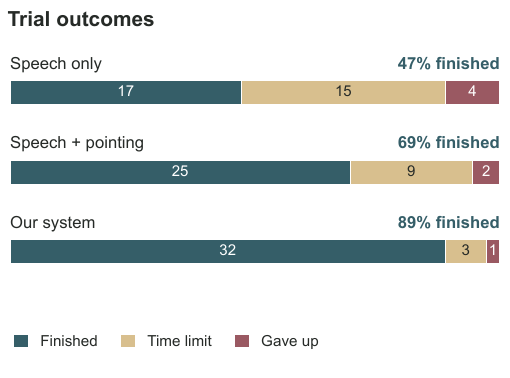}
\else
\includegraphics[width=\linewidth]{outcome.pdf}
\fi
\caption{Trial outcomes across all tasks (36 trials per condition). Segments show
counts; completion is participant-declared. The nominal limit was three
minutes; ``gave up'' denotes voluntary early termination.}
\Description{Three horizontal stacked bars, one per condition, with
counts inside the finished, time-limit, and gave-up segments. The finished
shares above the bars are 47, 69, and 89 percent for speech only, speech
with pointing, and our system.}
\label{fig:outcome}
\end{figure}

\begin{figure}[t]
\centering
\ifdefined\ReviewSubmission
\includegraphics[width=.68\linewidth]{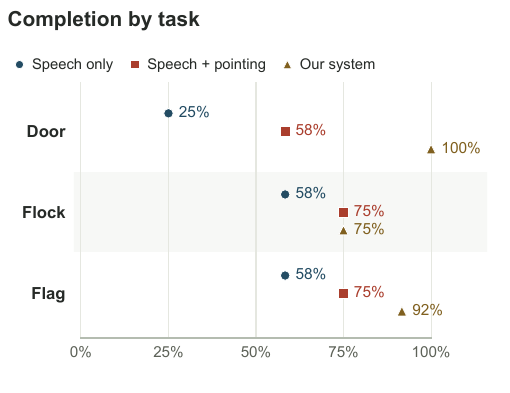}
\else
\includegraphics[width=\linewidth]{finished_by_task.pdf}
\fi
\caption{Participant-declared completion by task (12 trials per condition and task).}
\Description{Three task rows, Door, Flock, and Flag, with one labeled
point per condition on a shared zero-to-100-percent axis. The door rises
from 25 percent with speech alone to 100 percent with our system.}
\label{fig:outcome-task}
\end{figure}

\subsection{Task outcomes (RQ1)}

\subsubsection{Completion and trial outcomes}
\label{sec:results:completion}

Participants reported finishing 47\% of trials with speech only (17 of 36),
69\% with speech and pointing (25 of 36), and 89\% with our system
(32 of 36). Recorded overtime outcomes were 42\%, 25\%, and 8\%; abandoned
trials were 11\%, 6\%, and 3\%, respectively (Figure~\ref{fig:outcome}).

Participant completion proportions differed across conditions,
$Q(2)=14.25$, exact $p<.001$, $W=.59$. Our system exceeded speech alone
by 41.7 percentage points (95\% BCa CI [27.8, 61.1];
$p_{\mathrm{Holm}}=.006$, $r_{\mathrm{rb}}=1.00$): ten participants
completed more tasks and two tied. Neither pointing versus
speech ($p_{\mathrm{Holm}}=.109$) nor our system versus pointing
($p_{\mathrm{Holm}}=.125$; four participants improved, eight tied)
separated after correction. The supplement reports all pairwise tests.

The largest descriptive difference was on the door task: 3 of 12 finished
with speech alone and 12 of 12 with our system. The flag rose from 7
to 11 and the flock from 7 to 9 (Figure~\ref{fig:outcome-task}). These task
summaries identify where the observed completion difference was largest;
they do not constitute a separate interaction test or isolate a mechanism
by which the additional channels helped.

\begin{figure}[t]
\centering
\ifdefined\ReviewSubmission
\includegraphics[width=.68\linewidth]{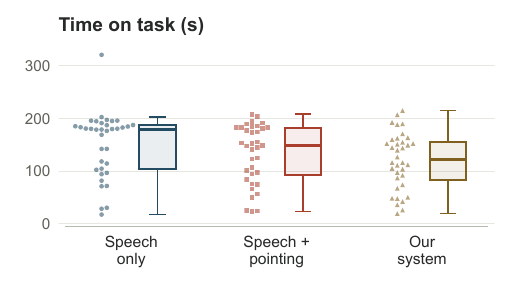}
\else
\includegraphics[width=\linewidth]{time_on_task.pdf}
\fi
\caption{Recorded duration from Start to logged closure, including unfinished trials
and overruns (36 per condition). Marks show trials; boxes show quartiles and
medians, with whiskers to extreme observations within 1.5 IQR of the
quartiles.}
\Description{Three conditions, each with 36 individual trial marks beside
a box plot, on a shared duration axis in seconds. Circles denote speech
only, squares speech with pointing, and triangles our system.}
\label{fig:time}
\end{figure}

\begin{figure}[t]
\centering
\ifdefined\ReviewSubmission
\includegraphics[width=.68\linewidth]{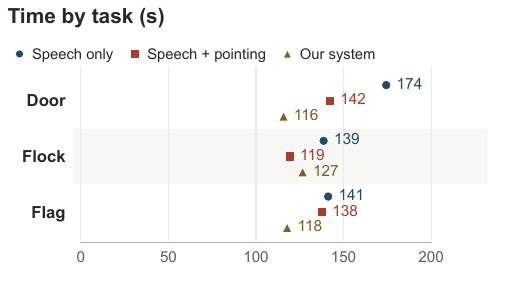}
\else
\includegraphics[width=\linewidth]{time_by_task.pdf}
\fi
\caption{Mean recorded duration in seconds by task (12 trials per condition and task).}
\Description{Three task rows, Door, Flock, and Flag, with one labeled
mean-duration point per condition on a shared axis in seconds. The largest
difference appears on the door task.}
\label{fig:time-task}
\end{figure}

\begin{figure}[t]
\centering
\ifdefined\ReviewSubmission
\includegraphics[width=.68\linewidth]{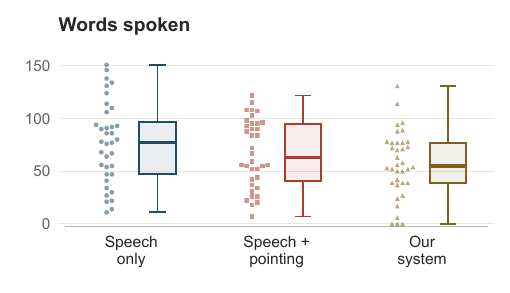}
\else
\includegraphics[width=\linewidth]{words_spoken.pdf}
\fi
\caption{Participant words per trial (36 trials per condition). Marks show trials; boxes
show quartiles and medians, with whiskers to extreme observations within 1.5
IQR of the quartiles.}
\Description{Three conditions, each with 36 individual trial marks beside
a box plot, on a shared axis of participant words per trial. Circles denote
speech only, squares speech with pointing, and triangles our system.}
\label{fig:words}
\end{figure}

\subsubsection{Trial duration and words spoken}
\label{sec:results:effort}

Recorded trial duration averaged 151~s with speech alone, 133~s with
pointing, and 120~s with our system. Mean participant word counts were
77, 67, and 57. Neither omnibus test detected a difference: duration
$Q(2)=2.17$, exact $p=.383$, $W=.09$; words $Q(2)=3.17$, exact $p=.249$,
$W=.13$ (Figures~\ref{fig:time} and~\ref{fig:words}).

These are durations and word counts across all recorded outcomes, not time
or words required to reach a successful result. Abandonment ends an
attempt early, while an overtime trial need not contain a finished result.
We therefore interpret these measures alongside completion and make no
claim of equal cost or reliably faster completion. The recorded durations
include overruns: 34 trials exceeded the nominal 180-second limit, so the
duration distribution is not truncated at that boundary.

By task, mean recorded door duration was 174~s with speech alone and 116~s
with our system. The corresponding differences for flock and flag
were 12~s and 23~s (Figure~\ref{fig:time-task}); these are descriptive
breakdowns of the same trials.

\begin{figure*}[t]
\centering
\includegraphics[width=\textwidth]{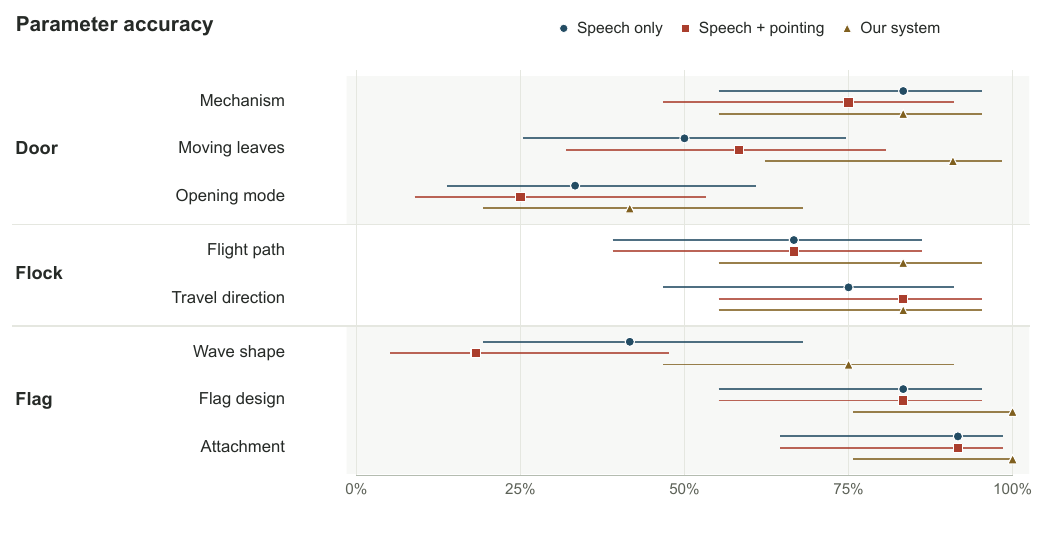}
\caption{Target matches among scorable trials (11--12 per condition and parameter),
with 95\% Wilson intervals.}
\Description{Eight parameters grouped into Door, Flock, and Flag tasks.
Each row has three points and horizontal 95 percent Wilson intervals.
Circles denote speech only, squares speech with pointing, and triangles
our system.}
\label{fig:accuracy}
\end{figure*}

\subsubsection{Target accuracy}
\label{sec:results:accuracy}

Participant-declared completion and target accuracy are separate outcomes.
The eight categorical accuracy measures cover all
scorable retained results, including unfinished trials (Figure~\ref{fig:accuracy}).
Six numerical measures also quantify how closely the authored parameters
match their targets. Complete accuracy
summaries, including denominators and paired tests, are retained in the
supplementary materials.

Our system had the highest observed match rate on six of the eight
categorical parameters and tied for highest on the remaining two
(Figure~\ref{fig:accuracy}). This descriptive advantage was consistent across
the plotted parameters, but no individual accuracy measure separated the
conditions after Holm adjustment across all fourteen measures. Flag wave shape matched in 5 of 12 speech-only,
2 of 11 pointing, and 9 of 12 trials with our system (paired $N=11$, exact
$p=.038$, adjusted $p=.488$). Our system also had lower mean wing-beat
rate error (2.6~Hz, compared with 6.0 and 6.4~Hz), but this exploratory
comparison did not survive adjustment (exact $p=.032$, adjusted $p=.446$).
Flock-count error gave exact $p=.111$, rather than evidence of a difference.

The Wilson intervals describe individual condition proportions and are not
tests of paired differences. For RQ1, our system had higher participant-declared completion than
speech alone; the accuracy, duration, and word-count comparisons did not
establish differences after the applicable corrections.

\subsection{Complementary input channels (RQ2)}

\subsubsection{Logged channel evidence}
\label{sec:division}

Pointing was present in 242 of 306 turns with speech and pointing (79\%)
and 253 of 312 turns with our system (81\%). In our system, span appeared
in 101 turns (32\%) and motion in 65 (21\%). Channels could coexist; these
percentages therefore do not partition turns. Span and motion were
unavailable in the other conditions.

The similar pointing rates suggest that participants continued to use
reference alongside the additional hand inputs. The records establish
channel presence, while the accounts below describe what participants
felt those channels helped express. They do not assign every output
parameter to a particular input channel.

\subsubsection{Recorded authoring exchanges}
\label{sec:results:exchanges}

Three selected door-task excerpts illustrate a multimodal request,
demonstration as a reply, and conversational refinement. These examples
show interaction patterns rather than their prevalence. Quotes are logged
speech or system replies; the supplement preserves the selected turn
ranges verbatim.

\emph{Combining words and movement.} P8 asked, ``Can you add a door here
that's hinged to the top and that opens like this?'' The turn contained
pointing, span, and motion evidence. The system replied, ``I made a
top-hinged door like that---like this?'' The request brought the object,
its location, and its intended movement into one utterance and
demonstration.

\emph{Answering through demonstration.} P11 said, ``I need a door right
here.'' The system asked, ``Show me with your hand how this door should
open.'' A demonstration-only reply led to a double-door proposal:
``I made it as double doors that swing open like that---like this?''
Movement served as the next conversational response. The demonstration-only
turn is represented by a generated log marker, not spoken words.

\emph{Refining an existing behavior.} P7 first requested a top-hinged door,
then asked it to swing outward. A subsequent turn combined speech and
motion: ``Can you animate it for me so it starts like this and swings
outward like this?'' The system replied, ``I've set the top hinge to swing
outward like that.'' Both refinement commands addressed the same door,
connecting verbal direction with a demonstrated opening movement.

\subsubsection{Participant accounts of channel use}
\label{sec:results:accounts}

The free-text answers were read together and grouped by what they were
about. Four themes organize the corpus, and we take them in turn.

\paragraph{Roles of hand input.}
Participants who chose our system described the hand as doing work
that words could not, and most named a specific class of intention it
carried. Quantities came up most often: P7 chose it because
``it was easier to describe measurements, heights and distances when able
to point and show,'' and returned to the same point about the birds task,
where the ability to show height and distance ``was extremely helpful.''
P14 described placing objects rather than naming them: ``I pointed and
showed the system where I wanted an object to be placed.'' P8 named the
property the paper is about directly: ``having the ability to show made it
easier to describe specific fine motions, which other modalities weren't
able to convey accurately.'' P11 gave the shortest version, that it ``feels
more natural to talk and use gestures,'' and P3 described the division of
labor as an ordering rather than a preference, saying that the three channels
together ``follow the common logics of what, where, how when building
something.''

\paragraph{Task-specific channel preferences.}
Eleven participants discussed whether a condition suited some tasks better
than others. P7 emphasized showing height and distance for the birds;
P3 favored our system for that task, whereas P6 preferred speech.
Preferences thus varied across participants for the same task.


\paragraph{Expressive scope.}
Of eleven participants answering the question about unexpressible
intentions, five reported none. The remaining responses concerned wingbeat
motion (two), size under time pressure (one), graphical changes outside
the task (one), and feedback or mutual understanding (two).


\paragraph{Interaction preferences.}
Two participants distinguished the appeal of the interaction technique
from their preference for a particular implementation. Three participants
selected a baseline in the forced choice. Preferences therefore concern
the complete configurations, while the accounts help explain what
participants valued about the available channels.


\subsection{Clarification responses (RQ3)}
\label{sec:grounding}

The system asked a clarifying question on 149 of 355 speech-only turns
(42\%), 123 of 306 pointing turns (40\%), and 127 of 312 turns with our system
(41\%). Of clarifications followed by another turn in the same trial,
nonverbal evidence occurred in 0 of 141 replies with speech alone, 89 of 115
with pointing (77\%), and 111 of 119 with our system (93\%). Under the
exclusive counting convention, the nonverbal replies with our system comprised 42
points, 27 spans, and 42 motions. Its clarification requests explicitly
expected showing 34 times and pointing 7 times; the remainder expected
speech or either.

These counts describe the channels participants used to answer
clarifications; they do not measure request resolution.


\subsection{Workload and preference}

\begin{figure*}[t]
\centering
\includegraphics[width=\linewidth]{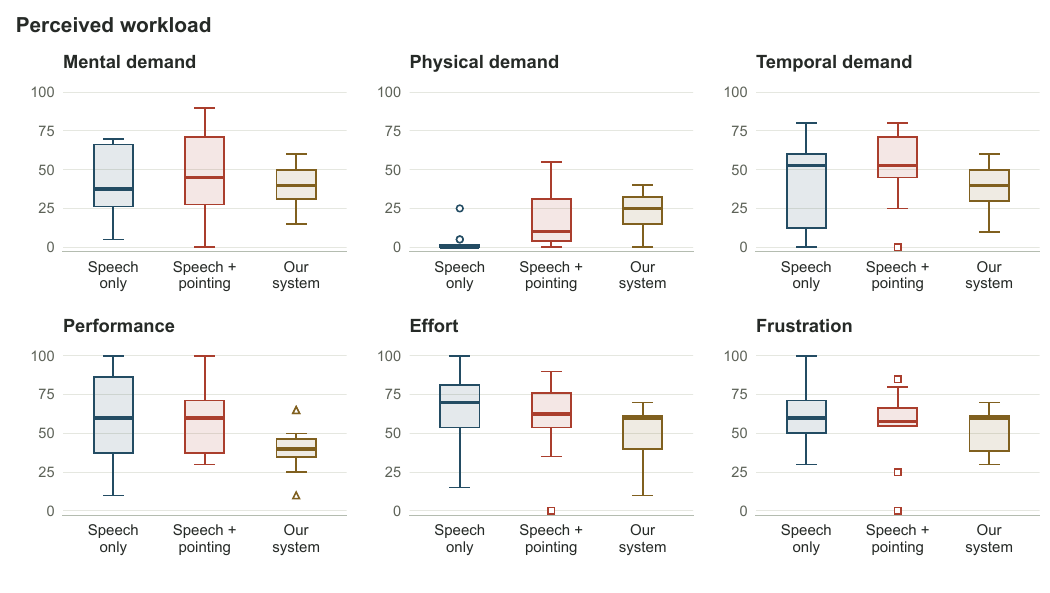}
\caption{NASA-TLX subscales (0--100; higher indicates greater workload, including
performance: 0 perfect, 100 failed). Boxes show quartiles and medians;
whiskers extend within 1.5 IQR of the quartiles, with outliers plotted
separately. Only physical demand differed after correction.}
\Description{Six panels of three box plots, one panel per TLX subscale.
Physical demand rises from near zero for speech only to a median of 25 for
our system; the other five subscales overlap heavily.}
\label{fig:tlx}
\end{figure*}

\subsubsection{Perceived workload}
\label{sec:results:workload}

Raw TLX means were 45.5, 48.8, and 40.8 for speech, pointing, and our system. The omnibus test did not detect a difference, $Q(2)=4.22$, exact
$p=.123$, $W=.18$. Physical demand did differ, with means 2.9, 18.8, and
23.3: $Q(2)=11.85$, exact $p=.002$, $W=.49$; Holm adjustment across the six
subscales gives $p=.009$. None of the other five subscales separated after
that adjustment (Figure~\ref{fig:tlx}).

Both hand conditions had higher physical demand than speech alone
($p_{\mathrm{Holm}}=.031$ for pointing and $.006$ for our system;
$r_{\mathrm{rb}}=1.00$ for both): seven and ten participants,
respectively, rated physical demand higher; the rest tied.
The comparison between our system and pointing
was not significant ($p_{\mathrm{Holm}}=.372$, $r_{\mathrm{rb}}=.33$).
Arm and hand tiredness, measured separately on a 0--10 scale, averaged 0.0, 1.2,
and 1.5; $Q(2)=10.80$, exact $p=.003$, $W=.45$. Our system exceeded
speech alone ($p_{\mathrm{Holm}}=.012$, $r_{\mathrm{rb}}=1.00$;
nine participants rated it higher, three tied).

These results identify a physical cost associated with hand input over the
short study tasks. The nonsignificant difference between the hand
conditions does not establish that adding span and demonstration is
cost-free, nor do the other null tests establish equal mental workload.
\subsubsection{Creativity support and experience}
The two CSI factors and eight additional experience ratings are reported
only in the supplementary summaries (Appendix~\ref{app:prompts}); no
significance claims are based on them.

\subsubsection{Rankings and overall preference}

None of the three ranking questions detected a difference across conditions. Asked which was easiest to work in, seven of the twelve put
our system first, four speech with pointing, and one speech alone
(mean ranks 1.67, 1.83 and 2.50). Asked which they would choose to use
again, the counts were six, five and one (1.67, 1.83 and 2.50). Asked
which best matched what they meant, six put speech with pointing first and
five our system, with one for speech alone, and the two mean ranks
tied at 1.75 against 2.50. None of the three omnibus tests detected a difference: $Q(2)=4.50$
to $4.67$, exact $p=.108$ to $.123$, $W\approx.19$.

Asked at the end of the session which of the three they would keep, nine
of the twelve chose our system, two chose speech with pointing, and
one chose speech alone. An exact multinomial test against equal choice probabilities, ordered by Pearson's statistic $\chi^2(2)=9.50$, gives $p=.012$.

The forced choice does not establish a separate pairwise preference
advantage over each baseline.

\section{Discussion}
\label{sec:discussion}

\subsection{Motion as conversational content}

The authoring loop makes demonstrated behavior available to later words
and demonstrations. Its value is that spatial and temporal detail can enter the
conversation without requiring the user to name every parameter. Higher
completion than speech alone supports this combined input configuration;
Section~\ref{sec:results:completion} reports the comparison.

Pointing remained frequent when span and motion were available. Together
with participant accounts (Section~\ref{sec:results:accounts}), this suggests
that demonstration can complement reference within the same conversation.
Establishing which channel supplies each parameter requires a separate
attribution analysis.

\subsection{Editable behavior after the demonstration}

An axis, pivot, range, and period make a demonstrated behavior addressable
by later instructions. The contribution is this shared representation for
speech and movement: an author can demonstrate an opening, inspect its
mechanism, and ask for a slower or wider version. Recordings and other
animation representations can also be edited; our choice exposes the
particular parameters needed for articulated behavior directly in the
conversation.

P7's refinement of an existing top-hinged door
(Section~\ref{sec:results:exchanges}) illustrates this continuity: verbal
direction and demonstrated motion address the same object. Later
instructions can refer to an existing behavior instead of requiring the
author to specify it again. Comparing parameter-based refinement with
editing recorded motion remains future work.


\subsection{Demonstration as a clarification response}

Clarification can request the form of evidence suited to an unresolved
detail. Participants supplied hand evidence in 93\% of replies to our
system's clarifications that had a subsequent turn in the same trial
(Section~\ref{sec:grounding}). This shows that movement can participate in
the answer, as P11's demonstration of a door opening illustrates.

Future studies could compare capture feedback and interpretation feedback
to understand how each supports refinement. The observed physical cost of
hand input also motivates studying when demonstration is worth the effort
and how that tradeoff changes during extended use.

%

\subsection{Demonstration in future agent architectures}
\label{sec:discussion:agents}

We propose that future agents built on large language models (LLMs) and
vision-language models (VLMs) should support demonstration as an explicit
form of instruction. Knowledge of how objects usually behave leaves open
how a particular user wants them to behave. ``Open the door'' admits many
plausible mechanisms and timings; a demonstration can supply the intended
choice. Combining speech and demonstration draws on complementary human
expressive abilities~\cite{mcneill1992hand,kita2003gesture,hostetter2011gestures},
letting users supply spatial and temporal intent without translating every
movement into verbal parameters.

Demonstration could connect \emph{perception, language, and action} around
user intent. Perception captures the movement's spatial and temporal
structure. Language and context establish its referent and instructional meaning.
Action produces behavior the user can inspect and revise, making the
outcome a shared reference for the next instruction.

Dubberly, Pangaro, and Haque frame conversation as interaction through which
participants coordinate goals and actions~\cite{dubberly2009interaction}.
Applied to future agents, this perspective suggests treating demonstrations,
questions, and previews as contributions to refining the intended behavior
across turns.

Shneiderman and Maes's interface-agent debate identifies user understanding
and control as central design concerns~\cite{shneiderman1997direct}.
Our authoring loop addresses them through visible previews and revision
of the inferred behavior.

PaLM-E provides a precedent for interleaving visual and continuous sensor
encodings with language~\cite{driess2023palme}; OpenVLA connects visual
representations and language to action prediction, using robot
demonstrations for training~\cite{kim2025openvla}. Together with co-speech
interfaces such as GesPrompt~\cite{hu2025gesprompt}, these works motivate
architectures in which a demonstration supplied during interaction becomes
an instruction that remains available across turns.

\paragraph{Preserve the relationship between words, movement, and intent.}
A demonstration should retain its temporal structure and alignment with
speech, the referenced object, and the current conversational turn.
A hand arc might specify a path, an opening range, or only a pace;
its instructional meaning depends on what the user is requesting or
answering. Future architectures could combine temporal motion embeddings
with explicit geometric summaries, preserving direction, rhythm, and
extent through interpretation. They should represent whether a movement
depicts desired behavior, selects a referent, or is incidental. The
agent then has evidence about both the movement and its communicative role.

\paragraph{Keep demonstrated behavior available for revision.}
Generative Agents retains and retrieves experiences to guide later
behavior~\cite{park2023generative}. For authoring, future work could similarly
link a demonstration to its target, spatial frame, and intended constraints.
A request such as ``the same path, but slower'' could then refer to that
earlier intent. Keeping the movement, inferred parameters, and preview
connected would support inspection and selective revision. This capability
could use model representations, external memory and tools, or a combination;
our parametric mechanisms illustrate one approach.

\paragraph{Let the agent request the appropriate form of evidence.}
An interaction policy could choose among requesting a name, a point, a
numerical value, or a demonstration according to the unresolved detail
and the user's available inputs, extending mixed-initiative
interaction~\cite{horvitz1999principles,cakmak2012questions}. Showing a
rhythm may require less verbal formulation than describing it. After
interpreting the answer, the agent should expose the proposed behavior
and support correction before committing consequential actions. Natural
interaction here means accommodating how users choose to communicate,
with alternatives to demonstration when another channel suits them better.

Future evaluations could test whether an agent follows different
demonstrations accompanying the same words, preserves intended constraints
during later edits, and reduces authoring effort. Animation blocking and
interactive teaching offer concrete settings: a designer might demonstrate
a wingbeat and revise its pace, while an instructor could show a hinge and
invite learners to vary its range. Our study motivates these possibilities;
their value across models and tasks remains to be evaluated.

\subsection{Scope and limitations}

\paragraph{A cumulative comparison in a small study.}
The twelve-person, within-subjects study compares three configurations of
one conversational system. It does not compare against conventional
keyframing, isolate motion from span, or establish general superiority in
animation authoring. Accuracy results are exploratory and do not establish
equivalence. No accuracy measure survived correction across the fourteen
measures; lower mean durations and word counts likewise did not establish
efficiency gains. Completion records participant satisfaction, not verified
target accuracy. Nine participants choosing our system is encouraging,
but does not establish a pairwise preference advantage over each baseline.
The retained results include unfinished trials. Condition and target-set
counterbalancing followed the schedule described in Section~\ref{sec:study}.
The study was not preregistered; its small sample and multiple exploratory
outcomes limit the precision and generalizability of the estimates.
The results concern the study implementation; subsequent updates to
clarification handling require separate evaluation.

\paragraph{One display and tracker.}
The flat display and hand tracker constrain the available depth cues and
workspace. A stereo or immersive replication would test how the
interaction changes when visual depth and hand position share a clearer
frame of reference. The recorded pointing channel also includes mouse
reference. Future instrumentation should distinguish input sources and
tracking availability.


\paragraph{A bounded behavior vocabulary.}
Hinge, slider, and spin mechanisms support a useful set of rigid motions;
flags use procedural cloth behaviors. General soft-body simulation,
event-conditioned behavior, reliable proportional edits, reference to
internal parts of composed props, and plural references remain outside or
at the edge of the supported vocabulary. Their absence limits what can be
authored regardless of the input channel.

\paragraph{Motion interpretation and confidence.}
The motion summaries depend on the captured segment: a fully retraced arc
can cancel the net angular sweep, and total travel alone does not determine
the range of a repeated motion. Evaluation should cover noise, occlusion,
and different cycle structures. Model-reported confidence has not
been calibrated against interpretation correctness; its threshold is a
design setting. Calibrating uncertainty and making inferred parameters
clearer would support more informed inspection and refinement.


\paragraph{Timing and model dependence.}
The recorded median fusion interval was 4.01~s; it excludes earlier speech
processing and therefore is not an end-to-end latency estimate. Trial
durations include logged overruns of the nominal three-minute task limit.
These measures must be interpreted alongside completion, rather than as
uncensored time to success. The hosted model configuration was held fixed
across conditions, but future service versions need not reproduce its
outputs exactly. Future measurements should separate speech capture,
transcription, reasoning, and execution, and record a consistent stopping
event for each trial. The current results cannot identify which stage
caused delays. Model and prompt snapshots and repeated-run evaluations
would help characterize output variability.

\paragraph{Participant reports and generalization.}
One coder organized the 101 main-study responses into themes. These
accounts describe experience rather than the prevalence of latent
constructs. Preferences concern the evaluated configurations. Longer use
and broader recruitment are needed to examine learning, accessibility,
and sustained physical effort. The comparison covers three short castle
tasks, mainly occasional 3D-tool users, and English interaction in a quiet
room. The lighting scene is a system example, not a separately evaluated
setting. Fixed task order and initial practice with all channels leave
transfer and learning unresolved. Expert workflows, open-ended authoring,
other languages, and users with different motor abilities require
evaluation.


\subsection{Future work}

Future work could test two-handed sizing and motion demonstration separately
to understand what each contributes. It could also examine which parts of
a request users express through speech, pointing, or movement. Following
clarification exchanges until users succeed or stop would show when asking
for a demonstration helps. Independent assessments of these exchanges
could distinguish receiving a new demonstration from resolving the problem.
A larger study, with its main comparisons and analysis plan set in advance,
could estimate the benefits more precisely. Comparing our system with
established animation tools would also show how it fits into broader
authoring workflows.

Future work could explore VR and autostereoscopic 3D displays to improve
depth judgment and calibration. VR places hands and geometry in a shared
3D frame; autostereoscopic displays add depth cues without worn glasses.
Both require evaluation of alignment, calibration effort, and viewing constraints.

Haptic gloves could provide tactile confirmation of contact and force
feedback during inspection of virtual parts, building on virtual-grasping
interfaces~\cite{hinchet2018dextres}. Lopes et al. used electrical muscle
stimulation to render resistance from virtual walls, gates, and
sliders~\cite{lopes2017haptics}. Such feedback could make a mechanism's
limits perceptible during preview. Comparisons with bare hands should
assess whether those benefits justify setup effort, fatigue, and
constraints on hand movement.

Other extensions include feedback for mechanism limits, reference to
several parts at once, and more expressive behavior constraints. Reliable proportional edits and
explicit correction of axis, pivot, range, and pace would make refinement
more predictable. Comparing captured-path feedback, motion previews, and
clarification policies separately would test their contribution to repair.
These changes should preserve the ability
to inspect and revise an inferred mechanism. A longitudinal deployment
could then examine whether familiarity changes channel selection and
whether physical effort remains acceptable during extended authoring.

\section{Conclusion}
\label{sec:conclusion}

``Like this'' can carry a specification. We have shown how this familiar
act of human communication can become part of conversational 3D authoring.
Speech, pointing, and demonstrated motion jointly specify behavior that
remains editable after the demonstration ends. A preview gives users something
to inspect and discuss; a clarification can invite another demonstration.
The contribution is this continuing relationship between showing an
intention and shaping its computational interpretation.

Across door, bird, and flag tasks, participants reported finishing 89\%
of trials with our system versus 47\% with speech alone; nine of twelve
preferred our system. It also achieved the highest observed match rates
on six categorical accuracy measures and tied on two. These findings
support the combined input configuration and motivate further study of
how its channels contribute to effective authoring.

More broadly, demonstration offers a way to connect perception, language,
and action around human intent. Future agents could use a movement as a
shared reference: something to interpret in context, ask about, and carry
forward into later instructions. This points toward interfaces that
accommodate how people express an idea through words and movement, while
keeping the resulting behavior visible and revisable. The demonstration can end
while the conversation about what it means continues.


\begin{acks}

We thank Sony Computer Science Laboratories and Sony Corporation for their generous support of this project. This work was also supported by the Japan Science and Technology Agency (JST) Moonshot R\&D Program under Grant Number JPMJMS2012. We thank our study participants for their time and perspectives, and the anonymous reviewers for their thoughtful and constructive feedback.

\end{acks}

\bibliographystyle{ACM-Reference-Format}
\bibliography{references}

\appendix

\section{Participant background}
\label{app:participants}

Table~\ref{tab:participants} reports the twelve main-study participants'
backgrounds; Section~\ref{sec:participants} summarizes the sample.
All were comfortable conversing in English: ten selected ``completely''
and two ``mostly.''

\begin{table*}[t]
\caption{Participant backgrounds, grouped by demographics, technology experience,
and communication habits. Pilots P1 and P2 are excluded.}
\label{tab:participants}
\footnotesize
\centering
\setlength{\tabcolsep}{3pt}
\begin{tabular*}{\textwidth}{@{\extracolsep{\fill}}lllllllll@{}}
\toprule
& \multicolumn{3}{c}{\textbf{Demographics}} & \multicolumn{3}{c}{\textbf{Technology experience}} & \multicolumn{2}{c}{\textbf{Communication habits}} \\
\cmidrule(lr){2-4}\cmidrule(lr){5-7}\cmidrule(l){8-9}
ID & Age band & Gender & \shortstack[l]{Dominant\\hand} & 3D tools & \shortstack[l]{Voice\\assistants} & \shortstack[l]{Hand\\tracking} & \shortstack[l]{Hand motion\\in conversation} & \shortstack[l]{Motion in\\work/study} \\
\midrule
P3 & 18 to 24 & man & right & few per year & few per year & once or twice & a fair amount & regularly \\
P4 & 18 to 24 & woman & right & once or twice & monthly & a few times & a fair amount & occasionally \\
P5 & 18 to 24 & woman & right & few per year & daily & a few times & a fair amount & occasionally \\
P6 & 18 to 24 & woman & right & few per year & once or twice & once or twice & a fair amount & regularly \\
P7 & 45 to 54 & man & right & once or twice & once or twice & once or twice & a fair amount & occasionally \\
P8 & 18 to 24 & man & left & few per year & once or twice & never & a little & never \\
P9 & 35 to 44 & woman & right & once or twice & daily & a few times & a fair amount & never \\
P10 & 18 to 24 & man & right & few per year & once or twice & once or twice & a fair amount & never \\
P11 & 55 to 64 & man & left & weekly & daily & regularly & a fair amount & never \\
P12 & 18 to 24 & woman & right & once or twice & few per year & once or twice & a fair amount & never \\
P13 & 25 to 34 & woman & right & once or twice & monthly & a few times & a little & occasionally \\
P14 & 25 to 34 & man & right & never & monthly & once or twice & a little & never \\
\bottomrule
\end{tabular*}
\par\smallskip
\parbox{\textwidth}{\footnotesize
\textit{Reading the table.} Ages are bands, not exact ages. Technology columns
report prior use; ``few per year'' means a few times a year. Hand motion
reports movement during conversation; motion in work/study
is how often participants describe or demonstrate movement. All entries are
self-reports, not observed behavior.}
\end{table*}


\section{Detailed system architecture}
\label{app:architecture}

Figure~\ref{fig:full-architecture} shows the runtime components and message
flow. Python sends the model named parts, live scene states, and hand
summaries; meshes remain in Unity, which executes edits and animation.
Model calls delay decisions without blocking rendering. Mode flags and
execution priorities arbitrate hand input; deictic events carry the active
mode to support interpretation.

\begin{figure*}[p]
  \centering
  \resizebox{\textwidth}{!}{
\begin{tikzpicture}[nf]
\useasboundingbox (0,0) rectangle (17.5,18.95);
\tikzset{
 fa group/.style={draw=nfLine,line width=.65pt,fill=nfBg},
 fa box/.style={draw=nfLine,line width=.5pt,fill=white},
 fa head/.style={font=\nfText\bfseries,anchor=base west,inner sep=0pt},
 fa note/.style={font=\nfSmall,text=nfGrey,anchor=base west,inner sep=0pt},
 fa code/.style={font=\nfSmall\ttfamily,text=nfGrey,anchor=base west,inner sep=0pt},
 fa flow/.style={nf arrow,line width=.9pt},
}
\draw[fa group] (0,13.70) rectangle (17.5,18.70);
\node[nf title] at (.28,18.27) {Unity front end};
\node[nf small,anchor=base east] at (17.22,18.28) {Local process \quad sensing, geometry + rendering};
\foreach \a/\b in {.28/5.50,6.14/11.36,12.00/17.22}
 \draw[fa box] (\a,14.56) rectangle (\b,17.95);
\node[fa head,text=nfRed] at (.52,17.53) {Capture hand evidence};
\node[fa code] at (.52,17.10) {HandPointer};
\node[fa note] at (.52,16.70) {Fingertip contact: part / prop / surface};
\node[fa code,text=nfAmber] at (.52,16.18) {DeicticEventStream};
\node[fa note] at (.52,15.78) {Events + mode; rolling hand buffer};
\node[fa note] at (.52,15.38) {Fetch a clip or watch a new episode};
\node[fa note,text=nfAmber] at (.52,14.94) {Raw hand frames go to the service};

\node[fa head,text=nfTeal] at (6.38,17.53) {Execute scene commands};
\node[fa code] at (6.38,17.10) {SelectionManager + EditExecutor};
\node[fa note] at (6.38,16.70) {Weighted mesh edits + scene undo};
\node[fa code] at (6.38,16.18) {SuggestionSpawner + GenericProp};
\node[fa note] at (6.38,15.78) {Props, articulated joints + paths};
\node[fa note] at (6.38,15.38) {Update the live scene at frame rate};
\node[fa note,text=nfTeal] at (6.38,14.94) {Scene outcomes return to dialogue};

\node[fa head,text=nfBlue] at (12.24,17.53) {Show the interpretation};
\node[fa code] at (12.24,17.10) {GlowRelay + MotionRibbon};
\node[fa note] at (12.24,16.70) {Reference highlight + captured hand path};
\node[fa code] at (12.24,16.18) {ClarifyFlow + CandidatePanel};
\node[fa note] at (12.24,15.78) {Questions, choices + touch answers};
\node[fa code] at (12.24,15.38) {Ghost};
\node[fa note,text=nfBlue] at (12.24,14.94) {Ghosts for adds driven by motion};
\node[fa note] at (.52,14.07)
 {\textbf{SceneStateReporter}\quad Live named parts, props, transforms + joints; refresh after scene changes};

\draw[fa flow] (.35,13.70) -- (.35,11.18);
\draw[fa flow,draw=nfTeal] (17.15,11.18) -- (17.15,13.70);
\node[nf small,font=\nfSmall\bfseries] at (8.75,13.38) {Local WebSocket \quad JSON messages};
\draw[nf rule] (8.72,11.51) -- (8.72,13.05);
\node[fa head] at (.90,12.91) {Evidence, answers + execution reports};
\node[fa code] at (.90,12.45) {deictic \quad motion\_clip \quad scene\_state};
\node[fa code] at (.90,12.05) {clarify\_answer \quad candidate\_picked};
\node[fa code] at (.90,11.65) {edit\_applied \quad (command ID, outcome)};
\node[fa head,text=nfTeal] at (9.18,12.91) {Commands, requests + visible feedback};
\node[fa code] at (9.18,12.45) {ai \quad highlight \quad status \quad transcript};
\node[fa code] at (9.18,12.05) {get\_motion \quad watch\_motion \quad watch\_cancel};
\node[fa code] at (9.18,11.65) {preview\_commit \quad preview\_cancel};

\draw[fa group] (0,5.10) rectangle (17.5,11.18);
\node[nf title] at (.28,10.75) {Python service};
\node[nf small,anchor=base east] at (17.22,10.76) {Local process \quad speech, evidence + dialogue};
\foreach \a/\b in {.28/5.50,6.14/11.36,12.00/17.22}
 \draw[fa box] (\a,6.18) rectangle (\b,10.35);
\node[fa head,text=nfBlue] at (.52,9.94) {Speech input + output};
\node[fa code] at (.52,9.52) {audio.py \quad speak.py};
\draw[nf rule] (.52,9.27) -- (5.26,9.27);
\node[fa note] at (.52,8.95) {Capture audio + detect speech};
\node[fa note] at (.52,8.52) {Send each utterance for transcription};
\node[fa note] at (.52,8.09) {Keep its original capture timestamps};
\node[fa note] at (.52,7.61) {Synthesize + play spoken replies};
\node[fa note] at (.52,7.18) {Manage playback and microphone input};
\node[fa note,text=nfBlue] at (.52,6.69) {A transcript starts a fusion turn};

\node[fa head,text=nfAmber] at (6.38,9.94) {Context + motion interpretation};
\node[fa code] at (6.38,9.52) {server.py Session \quad fusion.py};
\draw[nf rule] (6.38,9.27) -- (11.12,9.27);
\node[fa note] at (6.38,8.95) {Scene state + recent dialogue};
\node[fa note] at (6.38,8.52) {Open question, choices + pending ghost};
\node[fa note] at (6.38,8.09) {Speech window or recent events};
\node[fa note] at (6.38,7.61) {Contact episodes + dwell};
\node[fa note] at (6.38,7.18) {Hand-path summary + articulation fit};
\node[fa note,text=nfAmber] at (6.38,6.69) {Axis, pivot, range + pace};

\node[fa head,text=nfBlue] at (12.24,9.94) {Orchestrate each turn};
\node[fa code] at (12.24,9.52) {fusion.py \quad fuse()};
\draw[nf rule] (12.24,9.27) -- (16.98,9.27);
\node[fa note] at (12.24,8.95) {Serialize turns + call the model};
\node[fa note] at (12.24,8.52) {Fetch missing motion; reason again};
\node[fa note] at (12.24,8.09) {Apply confidence guards by operation};
\node[fa note] at (12.24,7.61) {Route commands, choices + questions};
\node[fa note] at (12.24,7.18) {Manage watches, previews + receipts};
\node[fa note,text=nfBlue] at (12.24,6.69) {Scene feedback continues the dialogue};

\draw[nf arrow] (5.50,8.52) -- (6.14,8.52);
\draw[nf arrow] (11.36,8.52) -- (12.00,8.52);
\node[fa note,text=nfInk] at (.52,5.72)
 {\textbf{Turn entry points:}\quad speech, an answer by touch, or a completed watched demonstration};
\node[fa note] at (.52,5.33)
 {\textbf{One authoring loop:}\quad interpret the request, show its meaning, and refine the scene};

\draw[fa flow] (.35,5.10) -- (.35,3.35);
\draw[fa flow,draw=nfBlue] (17.15,3.35) -- (17.15,5.10);
\draw[nf rule,dashed,draw=nfGrey] (0,4.70) -- (17.5,4.70);
\node[nf small,fill=white,font=\nfSmall\bfseries] at (8.75,4.70) {Hosted-service boundary};
\node[fa note] at (.90,4.20) {Audio or compact scene / motion context};
\node[fa note] at (9.18,4.20) {Transcripts, structured replies, speech audio};
\node[nf small] at (8.75,3.74) {Scene and motion are summarized; mesh arrays are never sent};

\draw[fa group] (0,.68) rectangle (17.5,3.35);
\node[nf title] at (.28,2.94) {Hosted AI services};
\node[nf small,anchor=base east] at (17.22,2.95) {Called by the Python service};
\foreach \a/\b in {.28/5.50,6.14/11.36,12.00/17.22}
 \draw[fa box] (\a,1.12) rectangle (\b,2.49);
\node[fa head] at (.52,2.08) {Speech recognition};
\node[fa note] at (.52,1.62) {Speech segment $\rightarrow$ transcript};
\node[fa head] at (6.38,2.08) {Fusion reasoning};
\node[fa note] at (6.38,1.62) {Semantic context $\rightarrow$ strict JSON reply};
\node[fa head] at (12.24,2.08) {Speech synthesis};
\node[fa note] at (12.24,1.62) {Reply text $\rightarrow$ speech audio};

\node[nf small,anchor=base west] at (0,.18)
 {Meshes + animation stay in Unity; Python interprets motion and coordinates dialogue.};
\end{tikzpicture}}
  \caption{Runtime architecture. Unity handles hands, scene execution, and feedback;
Python combines speech, scene context, and motion for interpretation and
dialogue. Hosted services provide transcription, reasoning, and speech. Arrows
show bidirectional message flow.}
  \Description{Three stacked groups show Unity, Python, and hosted services. Unity handles
hand capture, scene execution, feedback, and state reporting. Over a local
WebSocket, evidence and execution reports go to Python; commands, capture
requests, and preview controls return. Python handles speech, context, motion
interpretation, and turn orchestration. Across the hosted-service boundary,
audio and semantic context go out; transcripts, structured replies, and speech
return. Meshes and animation remain in Unity.}
  \label{fig:full-architecture}
\end{figure*}

%
%
%
%
%
%
%

\section{AI models}

\label{app:params}

Three hosted OpenAI models have distinct roles. Speech recognition
(\texttt{gpt-transcribe}) converts participant speech into text. Multimodal
reasoning (\texttt{gpt-5.4-mini}) combines that text with scene and hand
evidence to produce commands and replies. Speech synthesis
(\texttt{gpt-4o-mini-tts}) converts written replies into spoken audio.
Voice detection (Silero VAD) and hand tracking run locally. Models and
settings were held constant across conditions and sessions; the detailed
configuration will accompany the GitHub release
(Appendix~\ref{app:prompts}).

\section{Code and supplementary materials}
\label{app:prompts}
The accompanying review supplement contains all fourteen accuracy summaries,
participant-level omnibus and pairwise tests, descriptive questionnaire
results, scoring definitions, and selected anonymized interaction traces.
It includes an analysis of the retained numerical tables and an implementation
audit with code references. These materials support inspection of the
reported evidence; the current code snapshot is not a verified archival
build of every study session. The system source code, complete model prompts,
detailed configuration, and study tasks will be shared on GitHub.

\clearpage



\end{document}